\documentclass[acmsmall,screen]{acmart}
\AtBeginDocument{%
  }
\usepackage{cite}
\usepackage[ruled, linesnumbered]{algorithm2e}

\usepackage{amsmath,amssymb,amsfonts}
\usepackage{url}
\usepackage{booktabs}
\usepackage{xurl}
\usepackage{caption}
\usepackage{subcaption}
\usepackage{multirow}
\usepackage{multicol}
\usepackage{threeparttable}
\usepackage{algorithmic}
\usepackage{graphicx}
\usepackage{textcomp}
\usepackage{xcolor}
\usepackage{color}
\definecolor{ForestGreen}{rgb}{0.13,0.55,0.13}
\usepackage{listings}
\usepackage{xspace}
\usepackage{wrapfig}
\usepackage{booktabs}
\usepackage{array}
\usepackage{enumitem}
\usepackage{tcolorbox}
\usepackage{bbm}

\newcommand{\add}[1]{\textcolor{black}{#1}}

\setcopyright{cc}
\setcctype{by}
\acmDOI{10.1145/3808101}
\acmYear{2026}
\acmJournal{PACMSE}
\acmVolume{3}
\acmNumber{FSE}
\acmArticle{FSE094}
\acmMonth{7}
\acmSubmissionID{fse26mainb-p79-p}
\received{2025-09-12}
\received[accepted]{2026-03-24}
\begin{document}

\title{Cascaded Code Editing: Large-Small Model Collaboration for Effective and Efficient Code Editing}

\author{Chaozheng Wang}
\orcid{0000-0002-3935-7328}
\affiliation{%
  \institution{The Chinese University of Hong Kong}
  \city{Hong Kong}
  \country{China}
}
\email{adf111178@gmail.com}

\author{Zezhou Yang}
\orcid{0009-0008-9092-3381}
\affiliation{%
  \institution{The University of Hong Kong}
  \city{Hong Kong}
  \country{China}
}
\email{zezhouyang@connect.hku.hk}

\author{Shuzheng Gao}
\orcid{0000-0002-8102-480X}
\affiliation{%
  \institution{The Chinese University of Hong Kong}
  \city{Hong Kong}
  \country{China}
}
\email{szgao23@cse.cuhk.edu.hk}

\author[C. Gao]{Cuiyun Gao}
\orcid{0000-0001-8513-6836}
\authornote{Cuiyun Gao is the corresponding author.}
\affiliation{%
  \institution{The Chinese University of Hong Kong}
  \city{Hong Kong}
  \country{China}
}
\email{cuiyungao@outlook.com}

\author{Zongjie Li}
\orcid{0000-0002-9897-4086}
\affiliation{%
  \institution{Hong Kong University of Science and Technology}
  \city{Hong Kong}
  \country{China}
}
\email{zligo@cse.ust.hk}

\author{Yichen Li}
\orcid{0009-0009-8370-644X}
\affiliation{%
  \institution{The Chinese University of Hong Kong}
  \city{Hong Kong}
  \country{China}
}
\email{ycli21@cse.cuhk.edu.hk}

\author{Ting Peng}
\orcid{0009-0003-6970-0857}
\affiliation{%
  \institution{Tencent}
  \city{Guangzhou}
  \country{China}
}
\email{sakurapeng@tencent.com}

\author{Hailiang Huang}
\orcid{0009-0004-0655-9398}
\affiliation{%
  \institution{Tencent}
  \city{Guangzhou}
  \country{China}
}
\email{eraserhuang@tencent.com}

\author{Yuetang Deng}
\orcid{0009-0003-7060-4109}
\affiliation{%
  \institution{Tencent}
  \city{Guangzhou}
  \country{China}
}
\email{yuetangdeng@tencent.com}

\author{Michael R. Lyu}
\orcid{0000-0002-3666-5798}
\affiliation{%
  \institution{The Chinese University of Hong Kong}
  \city{Hong Kong}
  \country{China}
}
\email{lyu@cse.cuhk.edu.hk}


\begin{abstract}
Code editing constitutes a fundamental practice in software development, wherein developers modify existing codebases according to natural language requirements. Accurate code editing necessitates a comprehensive understanding of both the existing codebase and the modification requirements.  Although large language models (LLMs) have demonstrated promising performance in code editing tasks, they suffer from substantial inefficiency by generating entire modified files that largely consist of unchanged code. While smaller models could potentially address this inefficiency, they typically lack the capacity to effectively comprehend long code contexts required for accurate editing. To ensure both effectiveness and efficiency, we propose to decompose code editing into a two-stage cascade: \textbf{edit sketch generation}, wherein a large model first produces concise sketches representing the requisite modifications (the more challenging phase), and \textbf{edit sketch application}, wherein a smaller model integrates these sketches into the original code to produce the final output edited code (the simpler phase). This cascaded design reduces the number of tokens generated by the large model, as the majority of the output is handled by the smaller, more efficient model, thereby enhancing overall efficiency. However, the effectiveness of this approach is constrained by current small models' limited capabilities in handling long-context scenarios and cross-file dependencies, which are essential for accurate sketch application in real-world codebases. To address these limitations and enhance smaller models' sketch application capabilities, we introduce the first large-scale sketch application dataset comprising over 100K training instances and 800M tokens, along with a human-evaluated benchmark, and propose specialized training strategies including curriculum-based long-context training and multi-file augmentation. Our comprehensive experiments demonstrate that our cascaded framework inherently reduces inference costs compared to direct editing with large models. Furthermore, combining large models with our fine-tuned smaller models can achieve even superior performance. For instance, on the Aider benchmark, employing DeepSeek R1 as the edit sketch generation model alongside a fine-tuned Qwen2.5 Coder 14B model for the application phase improves Pass@2 11.1\% compared to direct editing with DeepSeek R1 alone. Additionally, the cascaded approach reduces execution time and cost by 13\% and 19\%, respectively, demonstrating both performance gains and efficiency improvements.

\end{abstract}


\maketitle

\section{Introduction}
Code editing is a crucial task in software development, which involves modifying existing codebases according to natural language instructions provided by developers.  Recent studies have shown that a significant portion of software development effort--estimated at 60-80\%--is dedicated to maintenance activities, including updating code to fix bugs, adapt to new environments, or enhance features \citep{banker1998software, kaur2015review}. As software systems evolve over lifecycles that can span 15-20 years, continuous code editing is essential to maintain their functionality and relevance \citep{mili2002reusing}. In collaborative and agile development environments, code editing enables multiple developers to work seamlessly on shared codebases, responding rapidly to feedback or changing specifications. Given the critical role and frequent nature of this activity, improving the efficiency and accuracy of code editing processes has become increasingly important for modern software engineering practices \citep{tao2012software}.
\begin{figure}[t]
    \centering
    \includegraphics[width=0.85\textwidth]{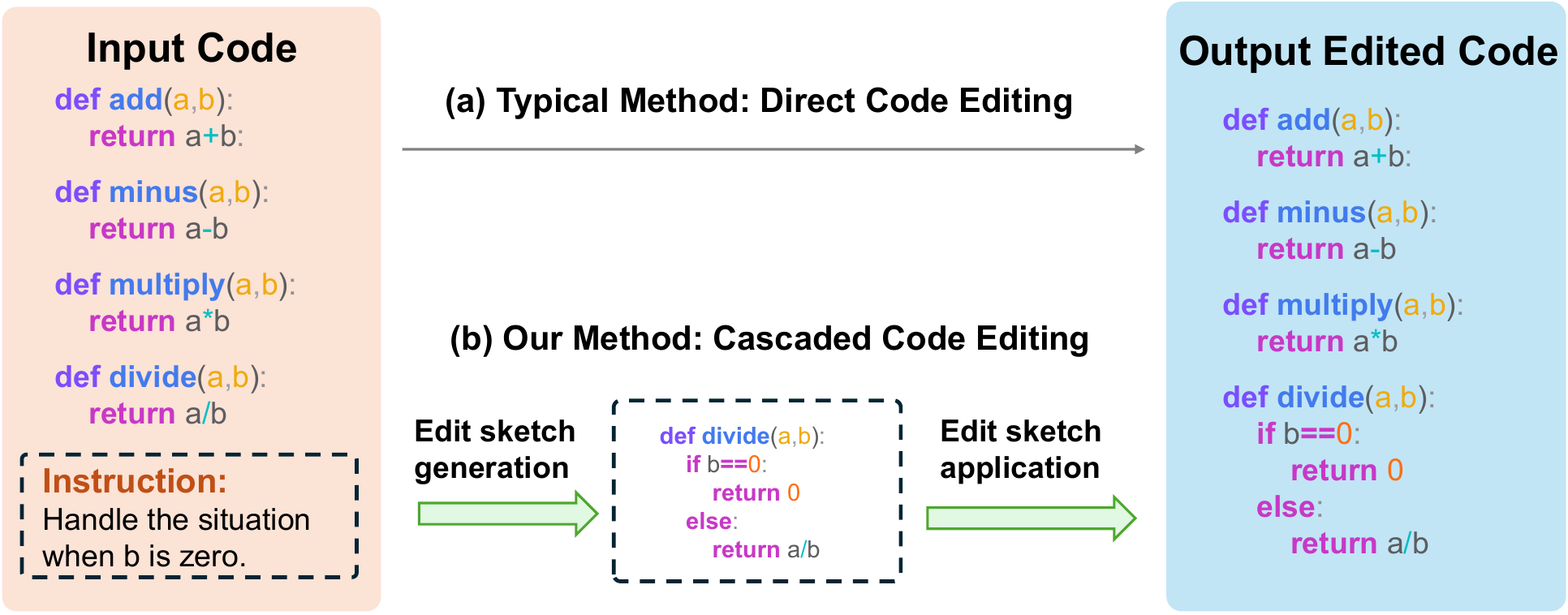}
    \vspace{-6pt}
    \caption{An example comparing direct editing (a) and cascaded code editing (b).}
    \label{fig:example_cascade}
    \vspace{-12pt}
\end{figure}

The emergence of Large Language Models (LLMs) has demonstrated remarkable capabilities across diverse software development tasks, including code completion~\citep{li2022cctest,wang2025beyond}, program repair~\citep{wong2025decllm,jelodar2025large}, and sophisticated code synthesis~\citep{ashrafi2025enhancing}. Building upon these advances, LLMs have enabled a new paradigm for automated code editing.
Guided by the natural language instructions for modifying the existing codebase, these models can generate targeted final source code (i.e., direct code editing). As shown in Figure \ref{fig:example_cascade} (a), after developers describe their requirements, the model produces the target code directly. This LLM-based code editing has emerged as a popular and valuable practice in contemporary AI-integrated development environments (AI-IDEs) such as Cursor \citep{cursor}, where developers can leverage these capabilities to substantially reduce the time and effort required for routine code modifications.

However, existing LLM-based methods struggle to balance effectiveness and efficiency in code editing tasks. On the one hand, the inherent complexity of modern codebases—characterized by intricate structures, cross-file dependencies, and files often exceeding several thousand tokens in length \citep{wang2024systematic, wang2025rag}—demands advanced reasoning capabilities. This necessitates the deployment of powerful large models, such as DeepSeek-V3 or Claude-4 \citep{liu2024deepseek, claude4}, to ensure reliable performance and accurate modifications. On the other hand, this reliance on advanced large models introduces significant efficiency bottlenecks. In real-world settings, regenerating entire modified files incurs high computational costs and slow inference speeds, leading to prolonged waiting times that can hinder developer productivity. As an illustration in our experiments (in Section \ref{sec:challeng1}), the average completion time for a single code editing request using DeepSeek R1 exceeds seven minutes in the real-world benchmark \citep{aider}. Moreover, code editing typically involves changes to only specific portions of the codebase, with empirical studies showing that an average of just 15\% of source code is altered during software updates, while the vast majority remains unchanged \citep{editrate}. This inefficiency compels large models to produce substantial redundant output that merely replicates the original code, wasting valuable computational resources and exacerbating latency issues.


To address the inefficiency issues of large models while ensuring the effectiveness, we present a cascaded code editing framework that decomposes the task into two complementary phases: \textbf{edit sketch generation} and \textbf{edit sketch application}, as shown in Figure \ref{fig:example_cascade} (b). In the edit sketch generation phase, a powerful but computationally expensive LLM analyzes the original code and natural language requirements to produce concise sketches—targeted code snippets representing the necessary modifications without regenerating unchanged portions of the codebase. In the edit sketch application phase, a smaller and more efficient specialized model merges these sketches into the original code to yield the final modified implementation. This decomposition leverages the superior capabilities of large models for the more challenging task of understanding complex codebases and modification requirements, while delegating the simpler application task to more efficient small models. By separating high-level edit planning from low-level implementation, this cascaded design maintains the quality advantages and simultaneously reduces the number of tokens generated by the large model, thereby enhancing overall efficiency.

        
Although the cascaded approach can enhance overall efficiency by offloading the bulk of code generation to smaller models, the edit sketch application task itself remains challenging. Current small models exhibit limited capabilities in handling long-context scenarios and cross-file understanding, which are essential for accurate sketch application in real-world codebases. These limitations prevent them from fully realizing the potential of high-quality edit sketches generated by large models.  To address the weak sketch application capability of current open-source models, we curate the first task-specific training dataset and an accompanying evaluation benchmark for this task. Specifically, we formalize sketch application as a prediction task: Given an original source file and a concise sketch that specifies targeted snippet-level edits while omitting unchanged code, the model must produce the final file with the edits correctly merged and no extraneous modifications.
The resulting corpus contains 118,280 training instances and 1,981 benchmark instances. Furthermore, we propose two progressive supervised fine-tuning (SFT) strategies to enhance sketch application performance. \textbf{Curriculum-based Long-Context SFT} (CLC SFT) employs staged training that begins with short-context instances for efficient convergence, then progresses to long-context data augmented with multi-file scenarios to enhance long-context and multi-file sketch application capabilities. \textbf{Generalized Curriculum Long-Context SFT} (G-CLC SFT) further incorporates general-domain coding tasks to improve out-of-domain generalization. Models trained with these approaches achieve average improvements of 7.1 and 40.9 points on CodeBLEU and exact match on our sketch application benchmark, respectively.

We further validate that our sketch application training yields substantial improvements in a practical cascaded code editing pipeline. For instance, when employing DeepSeek-R1 as the edit sketch generation model, our constructed dataset and progressive training strategies enhance the edit sketch application capabilities of Qwen2.5-Coder models ranging from 0.5B to 14B parameters by 12.6\% to 158.1\% in terms of Pass@2, compared to their untuned counterparts. Moreover, our cascaded approach has the potential to surpass direct code editing in performance. For instance, on the Aider benchmark, employing DeepSeek R1 as the edit sketch generation model alongside fine-tuned Qwen2.5-Coder 14B for the application phase improves Pass@2 by 11.1\%, compared to direct editing with DeepSeek R1 alone. This cascaded combination also reduces execution time by 13\% and computational costs by 19\%.  The results demonstrate that specializing the application phase can not only bridge but, in some cases, invert the effectiveness gap relative to direct editing of large-models, all while enabling a more efficient workflow.

In summary, this paper makes the following key contributions:

\begin{table}[h]
    \centering
    \caption{Effectiveness and efficiency of direct code editing on the Aider benchmark. Smaller models ($\leq$3B parameters) are typically offered for free by most API providers, while larger models incur usage costs based on input and output tokens.}
    \vspace{-6pt}
    \begin{tabular}{l|cc|llll}
    \toprule
     Models & Pass@1 & Pass@2 & \# Total Tokens & Throughputs & Time & Cost (\$) \\
     \midrule
    Qwen2.5-Coder 0.5B & 0.9 & 0.9 & 320.7K & 754.0 token/s & 0.12h & 0 \\
     Qwen2.5-Coder 1.5B & 2.7 & 3.6 & 286.9K & 320.8 token/s & 0.25h & 0 \\
     Qwen2.5-Coder 3B & 2.7 & 4.0 & 210.5K & 189.4 token/s & 0.31h & 0 \\
     Qwen2.5-Coder 7B & 2.7 & 4.9 & 321.7K & 146.3 token/s & 0.63h & 0.11 \\
     Qwen2.5-Coder 14B & 2.7 & 8.9 & 375.3K & 110.3 token/s & 0.89h & 0.17 \\
     \midrule
     DeepSeek-V3 & 29.3 & 56.0 & 357.1K & 26.4 token/s & 3.76h & 0.73 \\
     DeepSeek-R1 & 35.6 & 67.6 & 2.6M & 26.5 token/s & 27.31h & 6.54 \\
     \bottomrule
    \end{tabular}
    \label{tab:direct_editing_aider}
    \vspace{-6pt}
\end{table}

\begin{itemize}
    \item We propose a cascaded code editing framework that decomposes code editing into edit sketch generation and edit sketch application phases. This design optimizes the effectiveness-efficiency tradeoff by offloading simpler integration to smaller models, substantially reducing overall inference time and computational costs compared to direct editing approaches.

    \item We identify critical limitations in current open-source small models, particularly in long context handling and multi-file edit sketch application. To address these challenges, we introduce the first large-scale sketch application dataset with over 100K training instances and 800M tokens, accompanied by a human-evaluated benchmark and specialized progressive training strategies including CLC SFT and G-CLC SFT.

    \item Extensive experiments demonstrate that our enhanced small models, when integrated into the cascaded framework, achieve comparable or superior effectiveness compared to direct code editing with large models while maintaining substantially reduced inference time and computational costs. This advancement paves the way for more efficient AI-assisted software development workflows.
\end{itemize}

\section{Background and Motivation}\label{sec:bg}
This section provides the task formulation of code editing and analyzes two fundamental challenges that motivate our proposed cascaded framework. We first formalize the code editing problem and introduce the direct editing paradigm. Subsequently, we examine the effectiveness-efficiency trade-off inherent in direct code editing approaches, followed by an analysis of the limitations in current models' sketch application capabilities.

\subsection{Task Formulation}

Code editing is formally defined as modifying an existing codebase based on natural language instructions. Given an original source file (or set of files) $O$ and a natural language edit instruction $I$, the goal is to produce a modified file (or files) $F$ that incorporates the required changes while preserving the functionality and structure of unchanged portions.

This task encompasses various scenarios in software development, including bug fixes, feature additions, code refactoring, and optimization improvements. With the advancement of LLMs, it has become feasible to directly generate the edited code $F$ from the natural language instruction $I$ and original code $O$, enabling automated code editing at scale.

\begin{figure}[t]
    \centering
    \includegraphics[width=0.85\textwidth]{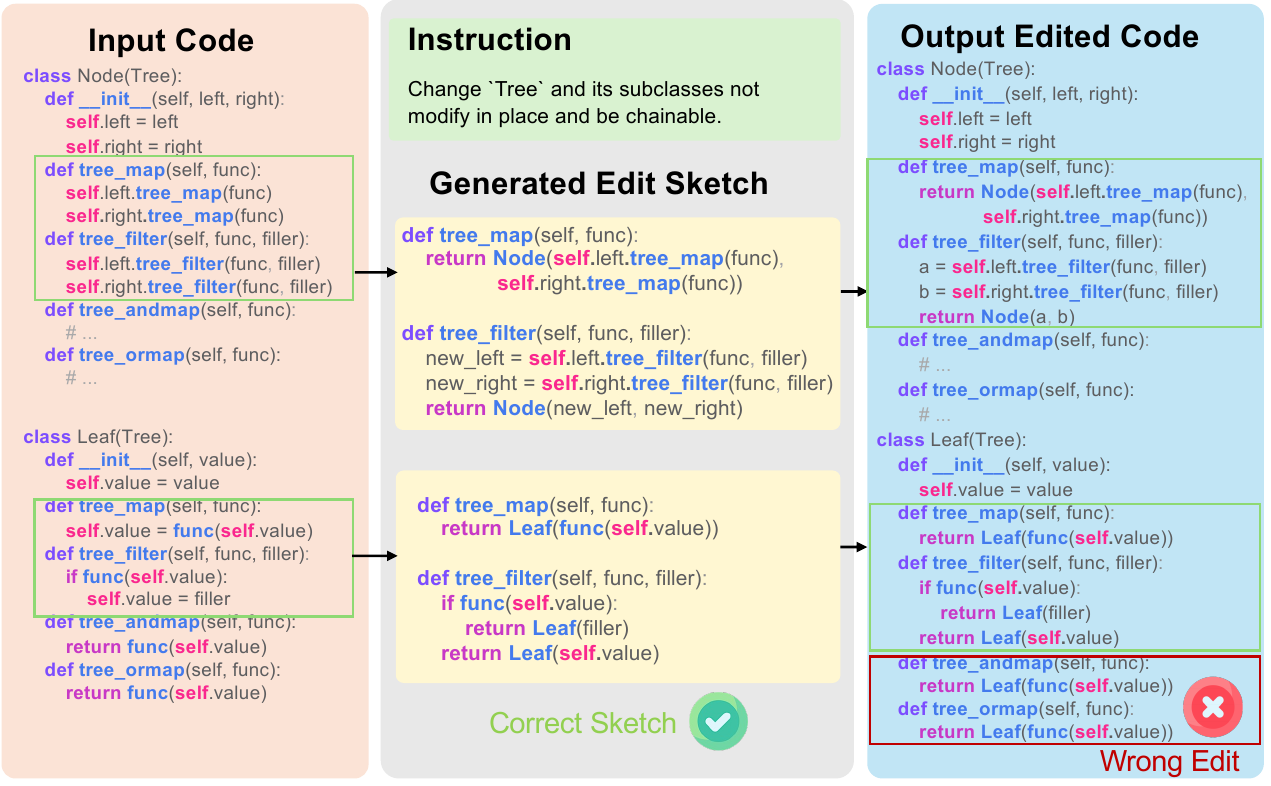}
    \vspace{-8pt}
   \caption{A motivating example about the challenge that small models face in precise sketch application.}
    \label{fig:example_fail}
    \vspace{-12pt}
\end{figure}
\subsection{Challenge 1: Effectiveness-Efficiency Trade-off in Direct Code Editing}\label{sec:challeng1}

Direct code editing faces a fundamental effectiveness-efficiency trade-off that limits its practical deployment. While large frontier models such as DeepSeek-V3 and DeepSeek-R1 achieve superior performance, they incur prohibitive computational costs. Conversely, smaller models offer efficiency but lack the capacity for accurate code editing. Table~\ref{tab:direct_editing_aider} presents performance and efficiency metrics on the Aider benchmark (detailed introduction of the benchmark and metrics can be referred to Section \ref{sec:benchmark}), clearly illustrating this challenge. The results reveal a pronounced performance gap between small/medium models and frontier large models. On the Aider benchmark, all small models perform poorly: even Qwen2.5-Coder 14B attains only a Pass@2 of 8.9, whereas DeepSeek-V3 achieves Pass@1/Pass@2 of 29.3/56.0, and DeepSeek-R1 further improves to 35.6/67.6 with its own stronger reasoning capabilities.
This indicates that model capacity and reasoning strength are pivotal for code editing quality, particularly on complex benchmarks where smaller models struggle to handle long, cross-file contexts and nuanced natural-language requirements.

However, achieving superior effectiveness comes at substantial computational expense. While smaller models demonstrate higher throughput (e.g., Qwen2.5-Coder 0.5B processes 754.0 tokens/s), the large models required for accurate editing operate at significantly lower speeds (DeepSeek-V3: 26.4 tokens/s, DeepSeek-R1: 26.5 tokens/s). More critically, the time and cost requirements become prohibitive: DeepSeek-V3 requires 3.76 hours and costs \$0.73 to complete the benchmark, while DeepSeek-R1, with its enhanced reasoning capabilities, demands an extraordinary 27.31 hours and \$6.54. This dramatic increase stems from DeepSeek-R1's verbose chain-of-thought generation, consuming 2.6M tokens compared to DeepSeek-V3's 357.1K tokens.

The fundamental inefficiency arises from the fact that large models must regenerate entire modified files, even when changes are localized to specific functions or modules. Empirical studies show that only 15\% of source code is typically altered during software updates~\citep{editrate}, while the vast majority remains unchanged. This forces sophisticated models to expend their computational capacity on reproducing unchanged code rather than focusing on the core editing logic, leading to suboptimal resource utilization and extended inference times that hinder developer productivity in practical deployment scenarios.

\subsection{Challenge 2: Limited Sketch Application Capabilities of Small Models}\label{sec:challenge2}

While the cascaded approach offers a promising solution to the effectiveness-efficiency trade-off by decomposing code editing into sketch generation and application phases, the edit sketch application task itself presents challenges. Current small models exhibit limitations in handling long-context scenarios and cross-file understanding, which are essential for accurate sketch application in real-world codebases.

While the cascaded framework enhances efficiency, its effectiveness is constrained by the limitations of current small models in the sketch application stage. Figure~\ref{fig:example_fail} illustrates a representative example: although the large model generates an accurate edit sketch in the first stage (e.g., redefining methods \texttt{tree\_map} and \texttt{tree\_filter} to be chainable), 
the small models fail to apply it correctly to the original code. Specifically, 
Qwen2.5-Coder 7B produces incomplete transformations, erroneously omitting or wrongly editing essential functions (\texttt{tree\_andmap} and \texttt{tree\_ormap}) from the final output despite correctly modifying the targeted methods.

The sketch application task, though simpler than sketch generation, still requires the model to precisely locate the segments in the original code that correspond to the sketch, apply modifications exclusively to those areas, and maintain consistency in long contexts--often spanning thousands of tokens or involving cross-file dependencies in real-world scenarios. Current open-source small models (typically 7B parameters or below) exhibit insufficient capabilities in these areas, leading to gaps in effectiveness compared to direct code editing approaches.
This limitation prevents small models from fully leveraging high-quality edit sketches, thereby undermining the potential of the cascaded approach to achieve both efficiency and effectiveness in practical code editing scenarios.

\section{Cascaded Code Editing with Enhanced Sketch Application}

To address the challenges identified in Sections \ref{sec:bg},
we propose a comprehensive solution consisting of: (1) a cascaded code editing framework that decomposes the task to leverage complementary model strengths (addressing Challenge 1), and (2) specialized dataset construction and training strategies to enhance smaller models' sketch application capabilities (addressing Challenge 2).

\subsection{Cascaded Code Editing Framework}
We introduce a cascaded code editing framework that decomposes the task into a guidance-and-implementation paradigm. As depicted in Figure \ref{fig:example_cascade} (b), code editing naturally lends itself to two stages: (1) \textbf{edit sketch generation}, where the model determines the locations and contents of required changes, producing a concise sketch \(S\) (e.g., targeted code snippets); and (2) \textbf{edit sketch application}, where the sketch is merged into the original code \(O\) to yield the final modified file \(F\).

The first stage is edit sketch generation, which requires a complex understanding of the codebase \(O\) and natural language instruction \(I\) to synthesize precise changes. This phase demands advanced capabilities typically found in large models to ensure correctness. The second stage, edit sketch application, is relatively simpler and focuses on mechanical integration: locating matching segments in \(O\), applying changes from \(S\), and maintaining syntactic consistency.

The key advantage of cascaded code editing lies in reducing the computational burden on large models while achieving overall efficiency gains. Taking the example in Figure~\ref{fig:example_cascade}, direct code editing requires a large model to generate the complete edited code using 50 tokens over 2.5 seconds. Our cascaded approach, however, decomposes this into two stages: the large model generates only a concise edit sketch using 20 tokens in 1 second, followed by a smaller model applying this sketch using 50 tokens in 0.2 seconds. While the total token consumption increases from 50 to 70 tokens, the overall inference time decreases significantly from 2.5 to 1.2 seconds, demonstrating the efficiency benefits of offloading bulk generation to faster small models.

\subsection{Dataset Construction}
We construct a specialized dataset tailored for evaluating and enhancing smaller models on the sketch-application task. This dataset combines real-world commit-based data with AI-assisted synthesized data, ensuring a diverse and high-quality corpus that covers various programming languages, edit complexities, and development scenarios. The construction process involves data collection as a foundational step, followed by targeted processing pipelines for commit data and synthesized data. We emphasize rigorous filtering mechanisms, including rule-based exclusions, AI-assisted validation, and consistency checks, to maintain data integrity and minimize noise.

\subsubsection{Data Collection}
The data collection phase serves as the bedrock for both commit-based and synthesized data pipelines, sourcing raw materials from diverse open-source GitHub repositories, as well as repositories from our cooperating institution, to ensure broad coverage. We begin by crawling a curated selection of popular repositories across 12 programming languages, including Python, JavaScript, Java, C++, and others. This process simultaneously aggregates historical commit records for the commit-based pipeline and collects corresponding source code files from the repositories, which are essential for the synthesized data pipeline as they provide the foundational codebases for AI-assisted edit generation.

We prioritize repositories with active development histories, such as those from Facebook, gRPC, and NVIDIA, to capture authentic code evolution patterns. \add{To ensure code quality and availability, we explicitly state the criteria for the 105 selected repositories, including strict license filtering (e.g., MIT/Apache 2.0) and popularity metrics (e.g., stars $>$ 1k).} In total, we collect over 342,883 commits from \add{these} 105 repositories, focusing on those involving substantive code modifications rather than trivial changes such as documentation updates. For synthesized data, the gathered source code files, approximately 1.3 million in total, ensure a mix of single-file and multi-file contexts. To enhance diversity, we mainly incorporate long-context examples (e.g., files exceeding 1,000 lines) and cross-file dependencies, simulating real-world editing scenarios. \add{Furthermore, the specific prompts used for synthesis and validation have been added to our anonymous repository to allow exact reproduction of the synthesized dataset.} All collected data undergoes initial preprocessing to remove duplicates and invalid entries.

\subsubsection{Commit Data Construction}
Building on the collected commit records, this pipeline transforms raw Git commits into structured triples of (original code, sketch, final code)  that conform to the sketch application task format. First, we extract diff blocks from each commit using Git's diff utility, converting them into concise sketches that highlight targeted changes without redundant unchanged code. For data quality assurance, only diffs containing at least 10 modification hunks are retained, while commits affecting solely non-code elements (e.g., comments, whitespace, or configuration files) are filtered out.

Following initial extraction, the commit data is refined by rule-based filtering, which involves discarding commits with pure additions/deletions of files, overly simplistic edits (e.g., single-line changes), or those dominated by formatting adjustments detected via static analysis tools such as clang-format \citep{clang}.  
Subsequently, AI-assisted validation leverages an LLM (e.g., DeepSeek-V3) to assess the quality of edit sketches, ensuring they are semantically meaningful and aligned with typical edit instructions. For instance, the LLM evaluates whether the diff represents a functional update, such as bug fixes or feature additions, rather than noise.

To further enhance uniqueness, we perform deduplication using simhash \citep{sadowski2007simhash} on the sketch snippets, removing entries with similarity exceeding 60\%—such as pairs involving a commit and its revert—to eliminate redundant or near-identical edits. This step typically discards approximately 20\% of the data instances. Finally, an AI consistency filter using DeepSeek-V3 verifies the integrity of each triple: the LLM simulates applying the sketch to the original code and checks if the result matches the final code from the commit. Inconsistent or erroneous samples are discarded. This process yields a high-quality commit-based subset of 73,762 samples, balanced across languages and edit types, providing authentic real-world examples for model training.
\subsubsection{Synthesized Data Construction}
To complement the commit-based data and expand coverage to underrepresented scenarios, we employ a pipeline that generates synthesized code edit examples. Specifically, we utilize a frontier LLM (e.g., DeepSeek-R1) 
to analyze each collected file by understanding its structure, functionality, and potential edit points. Based on this comprehension, the LLM generates a targeted sketch, simulating realistic edits such as adding features, refactoring code, or fixing hypothetical bugs. These sketches are designed to be concise and focus on substantive changes (e.g., logic alterations rather than cosmetic tweaks).

Subsequently, the LLM applies the generated sketch to the original code to produce the final code, creating a complete triple. We implement rule-based filtering to ensure data quality by excluding synthesized samples that contain only comments, format changes, or negligible modifications.
Specifically, this filtering employs regular expressions and heuristics to identify and remove trivial changes, such as whitespace-only modifications and comment-exclusive diffs.

Following rule-based filtering, we deploy a secondary validation process,
where another LLM (i.e., DeepSeek V3) reviews the sketch for coherence and applicability, ensuring it aligns with the file's context without introducing hallucinations. The final step mirrors the commit pipeline: the AI consistency check simulates the application process to confirm that the sketch snippet, when merged with the original code, precisely yields the final code. Any discrepancies lead to sample rejection. Finally, to ensure overall dataset uniqueness, the synthesized data is combined with the commit-based data and subjected to another round of deduplication using simhash, removing any cross-pipeline duplicates with similarity exceeding 60\%. This synthesis yields 44,518 additional samples, enriching the dataset with controlled variations in complexity (e.g., multi-file edits) and edge cases not prevalent in real commits, thus promoting better generalization during fine-tuning. 

\begin{table}[t]
    \centering
    \caption{Statistics of our dataset, where the tokenizer of Qwen2.5 Coder calculates the number of tokens.}
    \vspace{-6pt}
    \resizebox{\linewidth}{!}{
    \begin{tabular}{c|llllll}
    \toprule
     Split & \#Syn. Instance & \# Commit Instance & Avg. Code Len. & Avg. Sketch Len. &Avg. Final Len. & \#Tokens \\
     \midrule
     Train &44,518 & 73,762 & \multirow{2}{*}{3,230.0} & \multirow{2}{*}{293.5} & \multirow{2}{*}{3,366.9} & \multirow{2}{*}{827.4M}\\
     Test & 1,523 & 458 & & &  & \\
     \bottomrule
    \end{tabular}
    }
    \label{tab:statis}
    \vspace{-6pt}
\end{table}

\subsubsection{Benchmark Construction and Human Evaluation}\label{sec:bench}
To ensure the reliability of our evaluation and avoid potential leakage, we construct a separate benchmark using repositories distinct from those employed in dataset construction. This benchmark follows the same rigorous pipelines as described for commit-based and synthesized data, generating 2,168 samples. By sourcing from independent repositories, we maintain separation and promote fair assessment of model generalization on the sketch-application task.

Following data generation, two authors independently conduct human evaluation on the candidate samples. Each sample is scrutinized for common errors, including omissions (e.g., missed modifications), under-edits (e.g., incomplete changes), and subtle inaccuracies (e.g., syntactic inconsistencies or semantic mismatches) in the edit sketch application. Samples exhibiting these issues are discarded to uphold benchmark quality. After the initial independent assessments, the authors perform secondary confirmation by reviewing each other's evaluations and discussing any discrepancies.
This manual curation process results in a final benchmark comprising 1,981 high-fidelity samples, meaning that more than 90\% (1981/2168) of samples pass human evaluation, demonstrating the high quality of our automated pipeline for generating sketch-application data.

Detailed statistics of our datasets are presented in Table~\ref{tab:statis}. The training split comprises 44,518 synthesized instances and 73,762 commit-based instances and the test split includes 1,523 synthesized instances and 458 commit-based instances, enabling thorough evaluation of model performance on unseen data. The average lengths of our dataset are: 3,230.0 tokens for the original code, 293.5 tokens for sketches, and 3,366.9 tokens for the final code, totaling 827.4 million tokens to support comprehensive fine-tuning. 

\subsection{Proposed Training Strategies}
To optimize the fine-tuning of open-source models for the sketch-application task, we propose two progressive training strategies. Each builds upon supervised fine-tuning (SFT), incorporating adaptations to address long-context limitations, multi-file handling, and generalization concerns.

\subsubsection{Curriculum-based Long-Context SFT (CLC SFT)}

Standard supervised fine-tuning (SFT) typically employs truncation limits around 8,192 tokens to balance training efficiency ~\citep{muennighoff2023octopack, huang2024opencoder} with model performance, as longer contexts exponentially increase memory demands. However, this approach faces a fundamental trade-off: while 8,192 tokens suffice for most code editing scenarios, many real-world codebases contain files that exceed this limit, and truncating such instances can degrade the model's ability to handle extended code sequences effectively. Extending the truncation window to 16,384 tokens across the entire dataset would better preserve long-context information but introduces significant computational overhead. Training with such extended contexts requires memory optimization strategies like DeepSpeed ZeRO-3 with CPU offloading~\citep{ren2021zero}, leading to substantial training slowdowns when applied uniformly to all data.

To address this efficiency-capability trade-off, we propose Curriculum-based Long-Context SFT, a staged training approach that strategically allocates computational resources based on context length requirements. We partition our dataset based on the original code length, using 4,096 tokens as the threshold to create short-context (\(<4,096\) tokens) and long-context (\(\geq4,096\) tokens) subsets. This curriculum consists of two stages:

\textbf{Stage 1: Short-Context Foundation.} We fine-tune the model on short-context instances using an 8,192-token truncation window. This stage aims to establish baseline editing capabilities before handling longer contexts.

\textbf{Stage 2: Long-Context Enhancement.} We construct an enhanced training set combining: (1) original long-context instances from our dataset, and (2) artificially constructed multi-file instances created by randomly combining multiple short sketch-applying pairs. This multi-file augmentation serves dual purposes: increasing the volume of long-context training data and simulating real-world development scenarios where developers apply multiple modifications across different files within a single session. The model is then trained on this combined dataset with a 16,384-token window.

Formally, let \(\mathbf{x} = [\mathbf{o}; \mathbf{s}]\) denote the concatenated input sequence, where \(\mathbf{o}\) is the tokenized original code and \(\mathbf{s}\) is the edit sketch. The target output \(\mathbf{y}\) is the tokenized final code. The training objective minimizes the negative log-likelihood loss:
$\mathcal{L} = -\sum_{t=1}^{|\mathbf{y}|} \log P(y_t \mid \mathbf{x}, \mathbf{y}_{<t}; \theta)$,
where \(P(\cdot \mid \cdot; \theta)\) is the conditional probability distribution determined by model parameters \(\theta\).

This curriculum learning approach enhances long-context and multi-file editing capabilities while maintaining training efficiency through progressive complexity scaling.

\subsubsection{Generalized Curriculum Long-Context SFT (G-CLC SFT)}
Although CLC SFT effectively trains models for sketch application, focusing exclusively on this task may limit exposure to diverse editing patterns. In practical scenarios, sketch application involves complex interactions with varied code structures and contexts, requiring models to maintain broader editing capabilities beyond the specific sketch application task.

To mitigate this risk, we propose Generalized Curriculum Long-Context SFT, which integrates general-domain data from the OpenCoder SFT datasets~\citep{huang2024opencoder}. This external dataset encompasses a variety of code generation and question-answering tasks, providing a rich source of generic coding knowledge. We mix our task-specific data with the general-domain data at a 1:1 ratio during both curriculum stages, ensuring balanced exposure. This hybrid training preserves the model's universal capabilities while specializing in sketch application, leading to better robustness in complex, real-world code editing environments. This hybrid approach ensures models retain general coding competencies while acquiring specialized sketch application capabilities.

\section{Experimental Setup}

\subsection{Research Questions}

To guide our empirical evaluation of the cascaded code editing framework, we address the following key research questions, focusing on effectiveness, efficiency, and generalizability.

\textbf{RQ1: How effective and efficient is our cascaded code editing framework with specialized training compared to direct code editing?} We assess the overall performance of our complete approach on two established benchmarks, examining both effectiveness metrics (e.g., Pass@1/Pass@2) and efficiency indicators (e.g., token usage, wall-clock time, and cost) to demonstrate the dual advantages of our method.

\textbf{RQ2: What is the impact of different training strategies on edit sketch application performance?} We conduct ablation studies on our curated benchmark to evaluate how various training strategies affect the quality of edit sketch application, demonstrating the effectiveness of our proposed training methodology.

\textbf{RQ3: How do different training strategies for edit sketch application impact overall code editing performance?} We conduct ablation studies by integrating application models trained with different strategies into our cascaded framework and evaluate their performance on Aider and CanItEdit benchmarks to understand how sketch application training choices affect code editing effectiveness.

\textbf{RQ4: How transferable is the cascaded framework with enhanced edit sketch application capabilities to other LLMs?} We test generalizability by pairing fine-tuned edit sketch application models with alternative frontier LLMs for edit sketch generation, including Qwen3-235B \citep{yang2025qwen3}, Qwen3-Coder-480B \citep{qwen3coder}, and GPT-OSS-120B \citep{gptoss}.

\subsection{Selected Large Language Models}
\label{sec:prelim:llms}

We differentiate between the large-model component that performs edit sketch generation (and serves as a direct editing baseline) and the efficient small-model component that performs sketch application in the cascaded pipeline.

\textbf{Large-model component: DeepSeek-V3 and DeepSeek-R1.}
DeepSeek-V3 (DS-V3)~\citep{liu2024deepseek} is a mixture-of-experts LLM with 671B parameters for general-purpose capabilities. DeepSeek-R1 (DS-R1)~\citep{guo2025deepseek} is a follow-up to DS-V3 with enhanced reasoning via reinforcement learning. We use each of DS-V3 and DS-R1 in two roles: (i) as strong direct code-editing baselines, and (ii) as the edit sketch generation model in our cascaded framework.

\textbf{Small-model component: Qwen2.5-Coder (QC).}
For the edit sketch application (the second phase), we employ the Qwen2.5-Coder family~\citep{hui2024qwen2} across five sizes: 0.5B, 1.5B, 3B, 7B, and 14B. These models are chosen for their strong code competence, favorable efficiency–latency characteristics at small scales, and long-context support.

\subsection{Benchmarks and Evaluation Metrics}

To evaluate model performance in both direct and cascaded code editing, as well as the specific edit sketch application task, we utilize established benchmarks and tailored metrics as follows:

\subsubsection{Code Editing Benchmarks and Metrics}\label{sec:benchmark}

\textbf{Benchmarks.} We evaluate code editing performance on two established benchmarks including (1)
\textit{Aider Polyglot}~\citep{aider} is a comprehensive dataset containing 225 programming problems across multiple languages, simulating real-world repository code editing practices; and (2)
\textit{CanItEdit}~\citep{cassanocan} consists of 210 instructional code editing problems derived from 105 base problems, each with two instruction variants: ``lazy'' (minimal detail, mirroring typical user queries) and ``descriptive'' (detailed specifications). \add{To preclude data leakage, we quantify the similarity between our training corpus and the evaluation benchmarks. The analysis reveals minimal overlap, with average edit similarities of only 0.009 for Aider and 0.008 for CanItEdit, effectively demonstrating the distinctness of our evaluation set and the robustness of our model.}

\textbf{Effectiveness Metrics.} We primarily use Pass@$k$ to measure the percentage of problems solved correctly. Pass@1 measures the percentage of problems solved correctly on the first attempt, while Pass@2 allows a second attempt incorporating error feedback from the initial failure. For Aider Polyglot, we report both Pass@1 and Pass@2 following previous work \citep{qwen3coder,aggarwal2025nextcoder, seed2025seed}. For CanItEdit, we report Pass@1 for both lazy and descriptive NL descriptions \citep{seed2025seed, song2025seed}.

\textbf{Efficiency Metrics.} We employ several metrics to assess computational efficiency and practical deployment costs:
\begin{itemize}
    \item \textbf{Token usage}: For direct editing, we report the total tokens generated per benchmark (\# Total Tokens). For cascaded editing, we break it down into \# Sketch Tokens (generation stage) and \# Application Tokens (application stage).
    \item \textbf{Wall-clock time}: Total elapsed time to complete the benchmark, with stage-wise breakdowns (sketch generation time and application time) for cascaded approaches.
    \item \textbf{Cost}: Total monetary cost for API calls, calculated based on pricing for input and output tokens according to DeepInfra \citep{deepinfra}.
\end{itemize}

\subsubsection{Edit Sketch Application Benchmark and Metrics}
For the edit sketch application task, we employ the curated test set described in Section \ref{sec:bench} as the benchmark.

We assess performance using three complementary metrics: \textbf{CodeBLEU}~\citep{ren2020codebleu}, a code-oriented extension of BLEU that aggregates n-gram, keyword, abstract syntax tree (AST), and data-flow matching for lexical, structural, and semantic alignment, reported as a percentage (higher is better); \textbf{Fuzzy Match}, which checks for complete and correct application of sketches without omissions or extraneous changes, agnostic to exact positioning or formatting; and \textbf{Exact Match}, a binary metric requiring character-for-character identity with the reference final code.

\subsection{Implementation Details}

\begin{table}[t]
    \centering
    \caption{Hyper-parameter settings.}
    \vspace{-6pt}
    \resizebox{\linewidth}{!}{
    \begin{tabular}{c|cccccccc}
    \toprule
    Models   & Optimizer & LR &Scheduler & Batch Size & Max. Epochs& Precision &Train Seq. Len. & Max Gen. Len. \\
    \midrule
    Qwen2.5 Coder 0.5B  & \multirow{5}{*}{AdamW\citep{loshchilov2018decoupled}} & 5e-5 & \multirow{5}{*}{Cosine\citep{loshchilov2016sgdr}} & \multirow{5}{*}{32} & \multirow{5}{*}{3} & \multirow{5}{*}{BF16} &  &  \\
    Qwen2.5 Coder 1.5B & & 5e-5& & & & & 8,192 & 4,096\\
    Qwen2.5 Coder 3B & & 5e-5& & & & & $\sim$ &$\sim$ \\
    Qwen2.5 Coder 7B & & 1e-5& & & & & 16,384 & 8,192\\
    Qwen2.5 Coder 14B & & 1e-5& & & & & & \\
    \bottomrule
    \end{tabular}
    }
    \label{tab:hyper}
    \vspace{-8pt}
\end{table}

All the training experiments are run on a server with 8$\times$H20 GPUs, each with 96GB of graphic memory. The hyper-parameter settings for the tuning procedure are listed in Table~\ref{tab:hyper}. We enable the gradient checkpointing technique~\citep{chen2016training} to reduce GPU memory consumption across all models. For full-tuning experiments, we utilize ZeRO-3 techniques in DeepSpeed~\citep{ren2021zero, rajbhandari2020zero} to save GPU memory and improve training efficiency. Additionally, for long-context experiments involving models with 7B and 14B parameters under CLC SFT and G-CLC SFT, we enable CPU offloading in ZeRO-3~\citep{rajbhandari2020zero} to handle extended sequences effectively. We further incorporate Flash Attention~\citep{dao2022flashattention} to enhance training speed and optimize memory usage.

All inference experiments for large models, including DeepSeek-V3 and DeepSeek-R1, are deployed on a cluster with 16 H20 GPUs using the SGLang framework~\citep{zheng2024sglang} to support multi-machine deployment. For smaller models, such as the Qwen2.5-Coder series, we employ a single H20 GPU with vLLM~\citep{kwon2023efficient}, leveraging PagedAttention for efficient inference. Additionally, we incorporate the Flash-Attention technique~\citep{dao2022flashattention} for long-context optimization.

\section{Experiment Results}\label{sec:results}

\subsection{RQ1: Comparing Direct and Cascaded Code Editing}

\begin{table}[t]
    \centering
    \caption{Cascaded code editing with G-CLC SFT enhanced sketch application versus direct code editing on the Aider benchmark.}
    \vspace{-6pt}
    \resizebox{\textwidth}{!}{
    \begin{tabular}{l|cc|lllllll}
    \toprule
     Methods    &  Pass@1 & Pass@2 & \# Sketch Tokens & Sketch Time & \# Appl. Tokens & Appl. Time & \# Total Tokens & Total Time & Cost (\$)\\
     \midrule
     \multicolumn{10}{c}{DS-V3 as Edit Sketch Generation Model} \\
     \midrule
     DS-V3 (Direct) & 29.3 & \textbf{56.0} & - & - & - & - & 357.1K & 3.76h & 0.73 \\
     \midrule
     +QC 0.5B & 18.7 & 38.7 & 214.8K & 2.25h & 296.3K & 0.11h& 511.1K & 2.36h \color{ForestGreen}($\downarrow$36\%) & 0.46 \color{ForestGreen}($\downarrow$37\%) \\
    +QC 1.5B &24.0 & 48.4 & 221.6K & 2.32h & 302.5K & 0.26h & 524.1K & 2.58h \color{ForestGreen}($\downarrow$31\%) & 0.47 \color{ForestGreen}($\downarrow$36\%)\\
    +QC 3B &24.0 & 51.1 & 231.1K & 2.43h & 297.9K & 0.44h & 529.0K & 2.87h \color{ForestGreen}($\downarrow$24\%) & 0.48 \color{ForestGreen}($\downarrow$34\%)\\
    +QC 7B & 25.3 & 53.3  & 218.4K & 2.29h & 295.0K & 0.65h & 513.4K & 2.94h \color{ForestGreen}($\downarrow$22\%) & 0.54 \color{ForestGreen}($\downarrow$26\%)\\
    +QC 14B &\textbf{29.8} & 54.2 & 221.5K & 2.32h&301.2K& 0.82h & 522.7K & 3.14h \color{ForestGreen}($\downarrow$17\%) & 0.59 \color{ForestGreen}($\downarrow$19\%)\\
    \midrule
     \multicolumn{10}{c}{DS-R1 as Edit Sketch Generation Model} \\
     \midrule
     DS-R1 (Direct) & 35.6 & 67.6 & - & - & - & - & 2.6M & 27.31h & 6.54 \\
     \midrule
     +QC 0.5B & 20.4 & 41.3 & 2.1M & 22.23h & 302.8K & 0.11h& 2.4M & 22.34h \color{ForestGreen}($\downarrow$18\%) & 4.98 \color{ForestGreen}($\downarrow$24\%)\\
     +QC 1.5B &28.4 & 58.7 & 2.1M & 22.17h & 300.4K & 0.26h & 2.4M & 22.43h \color{ForestGreen}($\downarrow$18\%)& 4.96 \color{ForestGreen}($\downarrow$24\%)\\
     +QC 3B & 31.1 & 65.8 & 2.2M & 23.01h & 297.6K & 0.44h & 2.5M & 23.45h \color{ForestGreen}($\downarrow$14\%) & 5.22 \color{ForestGreen}($\downarrow$20\%)\\
     +QC 7B & 34.2 & 68.9 & 2.1M & 22.16h & 297.2K & 0.65h & 2.4M & 22.81h \color{ForestGreen}($\downarrow$16\%) & 5.03 \color{ForestGreen}($\downarrow$23\%)\\
     +QC 14B &\textbf{39.1} & \textbf{75.1} & 2.2M & 22.98h & 299.6K & 0.82h& 2.5M & 23.80h \color{ForestGreen}($\downarrow$13\%) & 5.32 \color{ForestGreen}($\downarrow$19\%)\\
     \bottomrule
    \end{tabular}
    }
    \label{tab:twostage_aider}
    \vspace{-10pt}
\end{table}

    

\begin{table}[t]
    \centering
    \caption{Cascaded code editing with G-CLC SFT enhanced sketch application versus direct code editing on the CanItEdit benchmark.}
    \vspace{-6pt}
    \resizebox{\textwidth}{!}{
    \begin{tabular}{l|cc|lllllll}
    \toprule
     Methods    &  Lazy & Descriptive & \# Sketch Tokens & Sketch Time & \# Appl. Tokens & Appl. Time & \# Total Tokens & Total Time & Cost (\$)\\
     \midrule
     \multicolumn{10}{c}{DS-V3 as Edit Sketch Generation Model} \\
     \midrule
     DS-V3 (Direct) & 58.1 & 66.7 & - & - & - & - & 84.6K & 0.92h & 0.10 \\
     \midrule
    +QC 0.5B & 51.4 & 57.1 & 37.5K & 0.39h&84.4K & 0.03h & 121.9K & 0.42h \color{ForestGreen}($\downarrow$54\%) & 0.06 \color{ForestGreen}($\downarrow$38\%) \\
    +QC 1.5B &53.3 & 63.8 & 37.8K & 0.39h & 80.9K &0.07h & 118.7K & 0.46h \color{ForestGreen}($\downarrow$50\%) & 0.06 \color{ForestGreen}($\downarrow$38\%)\\
    +QC 3B &58.1 & 66.7 & 35.8K& 0.38h& 83.4K &0.12h & 119.2K & 0.50h \color{ForestGreen}($\downarrow$46\%) & 0.06 \color{ForestGreen}($\downarrow$40\%)\\
    +QC 7B &\textbf{60.0} & \textbf{71.4} & 37.9K &0.39h &83.5K & 0.17h& 121.4K & 0.56h \color{ForestGreen}($\downarrow$39\%) & 0.07  \color{ForestGreen}($\downarrow$27\%)\\
    +QC 14B &59.0 & 70.5 & 37.5K & 0.39h&82.1K & 0.23h& 119.6K & 0.62h \color{ForestGreen}($\downarrow$33\%) & 0.09 \color{ForestGreen}($\downarrow$12\%)\\
    \midrule
     \multicolumn{10}{c}{DS-R1 as Edit Sketch Generation Model} \\
     \midrule
     DS-R1 (Direct) & 59.1 & \textbf{72.4} & - & - & - & - & 956.1K & 10.06h & 2.11 \\
     \midrule
     +QC 0.5B & 47.6 & 60.9 & 789.3K & 7.66h & 84.4K & 0.03h & 873.7K & 7.69h \color{ForestGreen}($\downarrow$24\%) & 1.75 \color{ForestGreen}($\downarrow$17\%)\\
    +QC 1.5B &56.2 & 62.9 & 811.5K & 7.87h & 85.1K & 0.07h & 896.6K & 7.94h \color{ForestGreen}($\downarrow$21\%) & 1.79 \color{ForestGreen}($\downarrow$15\%)\\
    +QC 3B &59.0 & 70.5 & 805.7K & 7.82h & 84.9K & 0.12h & 890.6K & 7.94h \color{ForestGreen}($\downarrow$21\%) & 1.78 \color{ForestGreen}($\downarrow$16\%)\\
    +QC 7B &60.9 & \textbf{72.4}& 809.1K &7.95h &84.5K & 0.17h& 893.6K & 8.12h \color{ForestGreen}($\downarrow$19\%) & 1.80 \color{ForestGreen}($\downarrow$15\%)\\
    +QC 14B &\textbf{61.9} & \textbf{72.4} & 808.6K & 7.94h & 83.2K & 0.23h & 891.8K & 8.17h \color{ForestGreen}($\downarrow$19\%) & 1.82 \color{ForestGreen}($\downarrow$14\%)\\
    \bottomrule
    \end{tabular}
    }
    \label{tab:twostage_can}
    \vspace{-10pt}
\end{table}

\textbf{Effectiveness}. Tables~\ref{tab:twostage_aider} and~\ref{tab:twostage_can} showcase the performance of the cascaded code editing approach using DeepSeek-V3 and DeepSeek-R1 as the edit sketch generation models, paired with specialized trained Qwen2.5-Coder models ranging from 0.5B to 14B as the edit sketch application models, on the Aider and CanItEdit benchmarks, respectively. We observe that the cascaded code editing approach achieves comparable or even superior performance to direct code editing methods. For instance, on the Aider benchmark, the combination of DeepSeek-V3 with trained Qwen2.5-Coder 14B achieves Pass@1 and Pass@2 scores of 29.8 and 54.2, which are comparable to the direct performance of DeepSeek-V3 (29.3/56.0). More remarkably, pairing DeepSeek-R1 with trained Qwen2.5-Coder 14B yields a Pass@2 performance of 75.1, representing an 11.1\% improvement over DeepSeek-R1's direct performance (67.6). This accuracy even surpasses SOTA closed-source LLMs Gemini-2.5-pro \citep{gemini} and Claude-4-Opus \citep{claude4} according to the official leaderboard \citep{leader}. Similar trends are observed on the CanItEdit benchmark, where our cascaded approach consistently matches or outperforms direct editing baselines. These findings demonstrate that large models can generate high-quality update sketches, and specialized trained small models can effectively apply these sketches to achieve superior performance.

Additionally, we find that the performance of the cascaded approach heavily depends on the size and capability of the edit sketch application model. Taking DeepSeek-V3 as an example, using trained Qwen2.5-Coder 0.5B compared to 14B results in performance gaps of 11.1 and 15.5 points on the Aider benchmark (Pass@1/Pass@2: 18.7/38.7 vs. 29.8/54.2), and 7.6 and 13.4 on the CanItEdit benchmark (Lazy/Descriptive Pass@1: 51.4/57.1 vs. 59.0/70.5), respectively. This indicates that while smaller models can benefit from specialized training, larger models are still needed to fully realize the potential of the high-quality sketches generated by large models.

\textbf{Efficiency}. Tables~4 and~5 present the efficiency metrics for cascaded code editing on the Aider and CanItEdit benchmarks, respectively. Compared to direct editing, the edit sketch generation stage in cascaded code editing, handled by the larger model, produces remarkably fewer tokens. For example, with DeepSeek-V3, the generated token count decreases by 38\% (from 357.1K to 221K on average) on the Aider benchmark and by 56\% (from 84.6K to 37.5K on average) on the CanItEdit benchmark. 
\add{Consequently, the execution time for Stage 1 is substantially lower than direct editing (e.g., reducing from 3.76h to $\sim$2.3h for DS-V3 on Aider), as the large model focuses solely on edit logic without reproducing the entire file.}
Although incorporating the sketch application results in a higher total token count overall (e.g., sketch + application tokens ranging from 510K to 522K on Aider and 118K to 119K on CanItEdit), the smaller models are more efficient in terms of throughput, leading to reduced total wall-clock time. 
\add{Regarding Stage 2, the runtime is highly correlated with the model size but remains a minor fraction of the total time. For instance, on Aider with DS-V3, the application time increases from 0.11h with the 0.5B model to 0.82h with the 14B model. Despite this increase, even the 14B model's application time is substantially less than the time saved in Stage 1, preserving the overall efficiency advantage.}


\subsection{RQ2: Impact of Training Strategies on Edit Sketch Application Performance}

\begin{table}
    \centering
    \caption{Performance of different small LLMs and training strategies in the edit sketch application task (CB: CodeBLEU; FM: Fuzzy Match; EM: Exact Match).}
    \vspace{-6pt}
    \resizebox{\linewidth}{!}{
    \begin{tabular}{l|cccc|cccc|cccc|cccc|cccc}
    \toprule
     \multirow{2}{*}{Metrics}  & \multicolumn{4}{c|}{QC 0.5B} & \multicolumn{4}{c|}{QC 1.5B} & \multicolumn{4}{c|}{QC 3B} & \multicolumn{4}{c|}{QC 7B} & \multicolumn{4}{c}{QC 14B} \\
         & Origin & + SFT & + CLC & + G-CLC & Origin & + SFT & + CLC & + G-CLC & Origin & + SFT & + CLC & + G-CLC & Origin & + SFT & + CLC & + G-CLC & Origin & + SFT & + CLC & + G-CLC \\
         \midrule
    \multicolumn{21}{c}{\textbf{Synthesized Data}} \\
    \midrule
    CB & 86.9 & 97.5 & \textbf{97.8} & 97.2 & 90.7 & \textbf{98.8} & \textbf{98.8} & 98.6 & 96.8 & \textbf{99.0} & 98.9 & 98.8 & 98.1 & \textbf{99.3} & \textbf{99.3} & 99.1 & 99.0 & \textbf{99.3} & \textbf{99.3} & 99.2 \\
    FM & 18.0 & 71.4 & \textbf{71.7} & 69.4 & 30.9 & 80.1 & \textbf{83.6} & 80.2 & 60.7 & 82.2 & \textbf{82.7} & 82.0 & 74.7 & 85.7 & \textbf{85.8} & 84.9 & 82.0 & 86.5 & \textbf{86.8} & 86.3 \\
    EM & 15.4 & 62.0 & \textbf{63.3} & 59.3 & 26.4 & 69.8 & \textbf{73.0} & 69.8 & 49.9 & \textbf{73.7} & 73.6 & 72.0 & 65.4 & 76.2 & \textbf{76.3} & 75.6 & 73.3 & 78.4 & \textbf{79.0} & 77.6 \\
    \midrule
    \multicolumn{21}{c}{\textbf{Commit Data}} \\
    \midrule
    CB & 75.5 & 95.2 & \textbf{95.4} & 95.6 & 79.8 & 98.3 & \textbf{98.5} & 97.8 & 92.9 & \textbf{98.8} & \textbf{98.8} & 98.4 & 95.2 & 98.9 & \textbf{99.0} & 98.7 & 98.3 & 99.0 & \textbf{99.2} & 98.8 \\
    FM & 0 & 40.7 & \textbf{44.9} & 41.0 & 3.9 & 67.0 & \textbf{73.4} & \textbf{70.6} & 18.2 & 73.2 & \textbf{76.0} & 72.3 & 36.3 & 80.6 & \textbf{82.3} & 81.0 & 55.9 & 82.9 & \textbf{86.1} & 83.3 \\
    EM & 0 & 30.1 & \textbf{33.0} & 32.8 & 2.8 & 51.9 & \textbf{56.0} & 54.9 & 11.8 & 56.9 & \textbf{60.7} & 59.6 & 26.3 & 63.1 & \textbf{65.4} & 64.7 & 40.2 & 66.3 & \textbf{67.0} & 66.2 \\
    \bottomrule
    \end{tabular}
    }
    \label{tab:apply}
    \vspace{-10pt}
\end{table}

Table \ref{tab:apply} presents an ablation study examining the impact of different training strategies on edit sketch application performance across current open-source small models (ranging from 0.5B to 14B parameters). We systematically compare our proposed CLC SFT and G-CLC SFT approaches against standard SFT with widely used settings (8,192 sequence length) \citep{huang2024opencoder, wei2023magicoder, wang2024exploring, yu2024wavecoder} to isolate the effects of different training methods. The results are divided into Synthesized Data and Commit Data subsets, evaluated using CodeBLEU (CB), Fuzzy Match (FM), and Exact Match (EM) metrics.

We observe a clear positive correlation between the edit sketch application capability of current open-source models and model size. For instance, Qwen2.5-Coder 14B achieves Exact Match scores of 73.3 on Synthesized Data and 40.2 on Commit Data. In contrast, these scores decline substantially for smaller models: the 1.5B variant records 26.4 and 2.8 (an average drop of 74\% relative to the 14B model), while the 0.5B variant yields 15.4 and 0 (an average drop of 86\%). Notably, the 0.5B model exhibits zero Fuzzy Match and Exact Match on Commit Data, underscoring the weak application abilities of current small-sized models (7B parameters and below). This trend highlights the limitations of smaller models in handling precise code merging, particularly in realistic commit scenarios involving complex, long-context codebases.

\begin{figure}[t]
    \centering
    \includegraphics[width=0.92\textwidth]{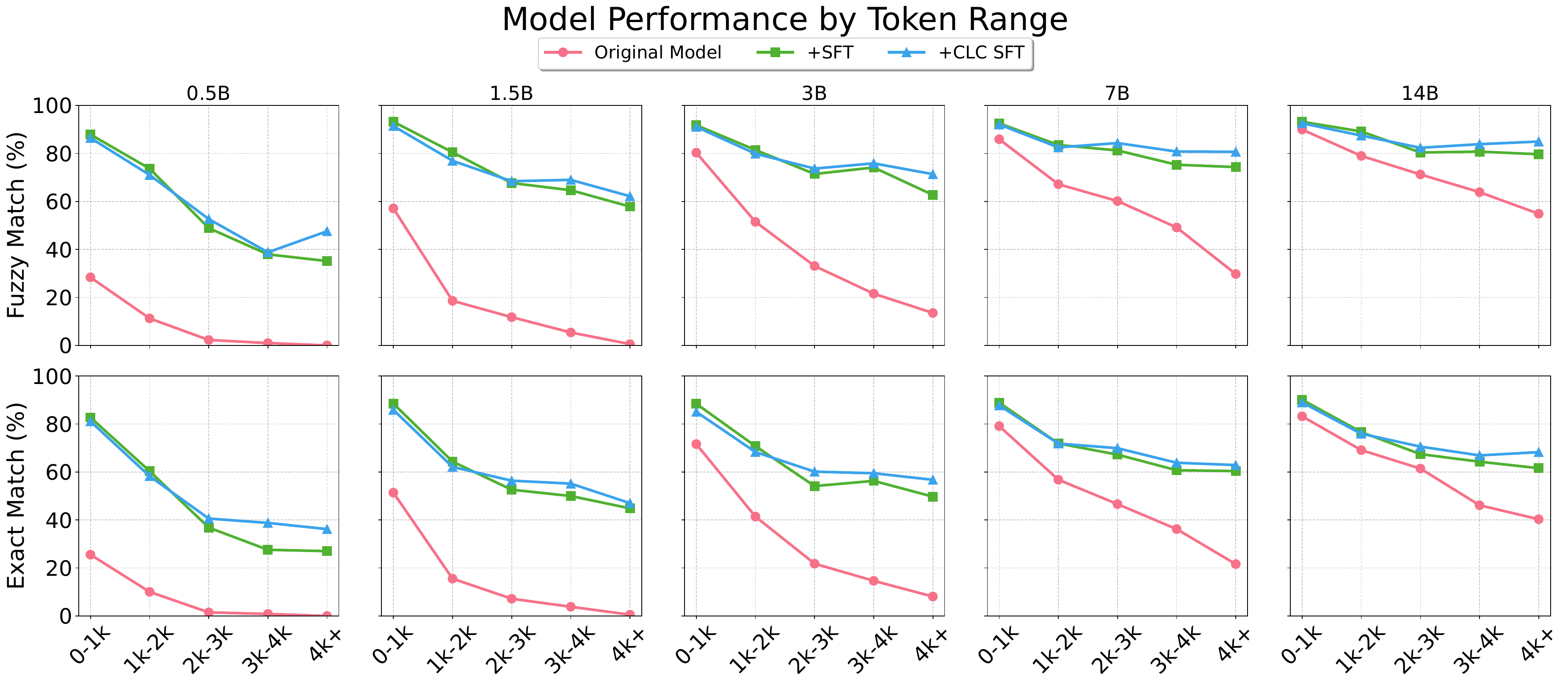}
    \vspace{-8pt}
    \caption{Model performance (FM and EM) on sketch application tasks by token range complexity. }
    \label{fig:token}
    \vspace{-8pt}
\end{figure}

Furthermore, we find that fine-tuning substantially enhances model performance on the edit sketch application task. Specifically, under the SFT setting, CodeBLEU, Fuzzy Match, and Exact Match improve by 0.3\%-12\%, 5.5\%-296\%, and 6.9\%-302\%, respectively, compared to the original models across the 0.5B to 14B range. These gains demonstrate that targeted data training can substantially bolster the application capabilities of current small models, with smaller models exhibiting more pronounced improvements.

In addition, Curriculum-based Long-Context SFT (CLC SFT) yields further performance enhancements, enabling models to progressively build long-context handling skills. For the 1.5B model on Commit Data, CLC SFT improves Fuzzy Match and Exact Match by 9.6\% and 7.9\% (relative to standard SFT), respectively. This indicates that our CLC training method which employs a curriculum learning approach to gradually improve sketch application and long-text capabilities. Figure \ref{fig:token} further illustrates the performance variations across five models ranging from 0.5B to 14B parameters, along with the original model, SFT, and CLC SFT versions, on different token ranges in the test set. The figure reveals that the original models exhibit substantial performance degradation as the length of the original code increases. For the more capable 14B model, Fuzzy Match and Exact Match performance degrade by 39\% and 52\%, respectively, when increasing the token length from 1K to 4K. The 0.5B and 1.5B models attenuate to 0\% in higher token ranges. However, the trained SFT and CLC SFT variants demonstrate noticeably reduced degradation as code length grows. Notably, CLC SFT, benefiting from specialized long-context training, exhibits reduced performance decay as token length increases from within 1k to over 4k tokens. For instance, in the 14B model, the relative decay on FM and EM metrics is 8.2\% and 23.3\%, respectively. The reductions decrease by 30.8\% and 28.3\% compared to the original Instruct model, and 6.3\% and 8.4\% compared to the SFT variant, underscoring its effectiveness in handling long-context sequences.

Finally, we note that Generalized Curriculum Long-Context SFT (G-CLC SFT) introduces a modest performance decline on this specialized edit sketch application benchmark. For instance, across models, Exact Match drops by 1\%-4\% relative to CLC SFT, attributable to the incorporation of general-domain tasks and data, which slightly dilutes task-specific proficiency. However, this trade-off enhances the model's generalization for practical use, as evidenced in downstream code editing pipelines (see Sections \ref{sec:rq3}).


\subsection{RQ3: Impact of Application Training on Cascaded Code Editing}\label{sec:rq3}

\begin{table}[th]
    \centering
    \caption{Performance of application-enhanced LCMs in the cascaded code editing task in the Aider benchmark.}
    \vspace{-6pt}
    \begin{subtable}[t]{0.48\textwidth}
    \centering
    \caption{Using DS-V3 as edit sketch generation model.}
    \resizebox{\linewidth}{!}{
        \begin{tabular}{l|ll}
    \toprule
    Methods & Pass@1 & Pass@2 \\
     \midrule
+QC 0.5B & 8.9 & 12.9 \\
+QC 0.5B w/ SFT & 10.2\color{ForestGreen}($\uparrow$14.6\%) & 23.1\color{ForestGreen}($\uparrow$79.1\%)\\
+QC 0.5B w/ CLC SFT & 15.2\color{ForestGreen}($\uparrow$70.8\%) & 30.2\color{ForestGreen}($\uparrow$134.1\%) \\
+QC 0.5B w/ G-CLC SFT & 18.7\color{ForestGreen}($\uparrow$110.1\%) & 38.7\color{ForestGreen}($\uparrow$200.0\%) \\
\midrule
+QC 1.5B &17.8 & 37.8 \\
+QC 1.5B w/ SFT & 22.7\color{ForestGreen}($\uparrow$27.5\%) & 47.6\color{ForestGreen}($\uparrow$25.9\%) \\
+QC 1.5B w/ CLC SFT & 23.1\color{ForestGreen}($\uparrow$29.8\%) & 48.0\color{ForestGreen}($\uparrow$27.0\%) \\
+QC 1.5B w/ G-CLC SFT & 24.0\color{ForestGreen}($\uparrow$34.8\%) & 48.4\color{ForestGreen}($\uparrow$28.0\%)\\
\midrule
+QC 3B &16.0 & 36.4 \\
+QC 3B w/ SFT &19.1\color{ForestGreen}($\uparrow$19.4\%) & 46.2\color{ForestGreen}($\uparrow$26.9\%) \\
+QC 3B w/ CLC SFT &23.1\color{ForestGreen}($\uparrow$44.4\%) & 50.2\color{ForestGreen}($\uparrow$37.9\%) \\
+QC 3B w/ G-CLC SFT &24.0\color{ForestGreen}($\uparrow$50.0\%) & 51.1\color{ForestGreen}($\uparrow$40.4\%) \\
\midrule
+QC 7B & 18.7 & 40.9 \\
+QC 7B w/ SFT & 22.7\color{ForestGreen}($\uparrow$21.4\%) & 47.6\color{ForestGreen}($\uparrow$16.4\%) \\
+QC 7B w/ CLC SFT & 24.0\color{ForestGreen}($\uparrow$28.3\%) & 51.1\color{ForestGreen}($\uparrow$24.9\%) \\
+QC 7B w/ G-CLC SFT & 25.3\color{ForestGreen}($\uparrow$35.3\%) & 53.3\color{ForestGreen}($\uparrow$30.3\%) \\
\midrule
+QC 14B &25.3 & 52.0 \\
+QC 14B w/ SFT & 26.2\color{ForestGreen}($\uparrow$3.6\%) & 52.0\color{gray}($=$0.0\%) \\
+QC 14B w/ CLC SFT & 27.1\color{ForestGreen}($\uparrow$7.1\%) & 52.4\color{ForestGreen}($\uparrow$0.8\%) \\
+QC 14B w/ G-CLC SFT & 29.8\color{ForestGreen}($\uparrow$17.8\%) & 54.2\color{ForestGreen}($\uparrow$4.2\%) \\
     \bottomrule
    \end{tabular}
    }
    \end{subtable}
    \hfill
    \begin{subtable}[t]{0.48\textwidth}
    \centering
    \caption{Using DS-R1 as edit sketch generation model.}
    \resizebox{\linewidth}{!}{
        \begin{tabular}{l|ll}
    \toprule
    Methods & Pass@1 & Pass@2 \\
     \midrule
+QC 0.5B & 9.8 & 16.0 \\
+QC 0.5B w/ SFT & 14.7\color{ForestGreen}($\uparrow$50.0\%) & 26.2\color{ForestGreen}($\uparrow$63.8\%)\\
+QC 0.5B w/ CLC SFT & 17.8\color{ForestGreen}($\uparrow$81.6\%) & 36.9\color{ForestGreen}($\uparrow$130.6\%) \\
+QC 0.5B w/ G-CLC SFT & 20.4\color{ForestGreen}($\uparrow$108.2\%) & 41.3\color{ForestGreen}($\uparrow$158.1\%) \\
\midrule
+QC 1.5B & 16.4 & 26.2 \\
+QC 1.5B w/ SFT & 22.7\color{ForestGreen}($\uparrow$38.4\%) & 47.6\color{ForestGreen}($\uparrow$81.7\%) \\
+QC 1.5B w/ CLC SFT & 27.6\color{ForestGreen}($\uparrow$68.3\%) & 57.3\color{ForestGreen}($\uparrow$118.7\%) \\
+QC 1.5B w/ G-CLC SFT & 28.4\color{ForestGreen}($\uparrow$73.2\%) & 58.7\color{ForestGreen}($\uparrow$124.0\%)\\
\midrule
+QC 3B & 16.4 & 41.3 \\
+QC 3B w/ SFT & 26.6\color{ForestGreen}($\uparrow$62.2\%) & 56.0\color{ForestGreen}($\uparrow$35.6\%)\\
+QC 3B w/ CLC SFT & 29.8\color{ForestGreen}($\uparrow$81.7\%) & 62.2\color{ForestGreen}($\uparrow$50.6\%) \\
+QC 3B w/ G-CLC SFT & 31.1\color{ForestGreen}($\uparrow$89.6\%) & 65.8\color{ForestGreen}($\uparrow$59.3\%) \\
\midrule
+QC 7B & 21.3 & 46.7 \\
+QC 7B w/ SFT & 30.2\color{ForestGreen}($\uparrow$41.8\%) & 60.9\color{ForestGreen}($\uparrow$30.4\%)\\
+QC 7B w/ CLC SFT & 35.1\color{ForestGreen}($\uparrow$64.8\%) & 65.8\color{ForestGreen}($\uparrow$40.9\%) \\
+QC 7B w/ G-CLC SFT & 34.2\color{ForestGreen}($\uparrow$60.6\%) & 68.9\color{ForestGreen}($\uparrow$47.5\%) \\
\midrule
+QC 14B & 38.3 & 66.7 \\
+QC 14B w/ SFT & 38.3\color{gray}($=$0.0\%) & 69.3\color{ForestGreen}($\uparrow$3.9\%) \\
+QC 14B w/ CLC SFT & 39.6\color{ForestGreen}($\uparrow$3.4\%) & 73.3\color{ForestGreen}($\uparrow$9.9\%) \\
+QC 14B w/ G-CLC SFT & 39.1\color{ForestGreen}($\uparrow$2.1\%) & 75.1\color{ForestGreen}($\uparrow$12.6\%) \\
     \bottomrule
    \end{tabular}
    }
    \end{subtable}
    \label{tab:aider_train}
    \vspace{-6pt}
\end{table}

\begin{table}[th]
    \centering
    \caption{Performance (Pass@1) of application-enhanced LCMs in the cascaded code editing task in the CanItEdit benchmark.}
    \vspace{-6pt}
    \begin{subtable}[t]{0.48\textwidth}
    \centering
    \caption{Using DS-V3 as edit sketch generation model.}
    \resizebox{\linewidth}{!}{
        \begin{tabular}{l|ll}
    \toprule
    Methods & Lazy & Descriptive \\
     \midrule
     +QC 0.5B & 20.9 & 28.6 \\
+QC 0.5B w/ SFT & 41.9\color{ForestGreen}($\uparrow$100.5\%) & 49.5\color{ForestGreen}($\uparrow$73.1\%)\\
+QC 0.5B w/ CLC SFT & 45.7\color{ForestGreen}($\uparrow$118.7\%) & 53.3\color{ForestGreen}($\uparrow$86.4\%) \\
+QC 0.5B w/ G-CLC SFT & 51.4\color{ForestGreen}($\uparrow$146.4\%) & 57.1\color{ForestGreen}($\uparrow$99.7\%) \\
\midrule
+QC 1.5B &36.2 & 42.9 \\
+QC 1.5B w/ SFT & 42.9\color{ForestGreen}($\uparrow$18.5\%) & 53.3\color{ForestGreen}($\uparrow$24.2\%) \\
+QC 1.5B w/ CLC SFT & 51.4\color{ForestGreen}($\uparrow$42.0\%) & 60.0\color{ForestGreen}($\uparrow$39.9\%) \\
+QC 1.5B w/ G-CLC SFT &53.3\color{ForestGreen}($\uparrow$47.2\%) & 63.8\color{ForestGreen}($\uparrow$48.7\%) \\
\midrule
+QC 3B &48.6 & 53.3 \\
+QC 3B w/ SFT & 53.3\color{ForestGreen}($\uparrow$9.7\%) & 60.0\color{ForestGreen}($\uparrow$12.6\%) \\
+QC 3B w/ CLC SFT & 55.2\color{ForestGreen}($\uparrow$13.6\%) & 64.8\color{ForestGreen}($\uparrow$21.6\%) \\
+QC 3B w/ G-CLC SFT & 58.1\color{ForestGreen}($\uparrow$19.5\%) & 66.7\color{ForestGreen}($\uparrow$25.1\%) \\
\midrule
+QC 7B & 55.2& 61.9 \\
+QC 7B w/ SFT & 58.1\color{ForestGreen}($\uparrow$5.3\%) & 62.9\color{ForestGreen}($\uparrow$1.6\%) \\
+QC 7B w/ CLC SFT & 59.0\color{ForestGreen}($\uparrow$6.9\%) & 69.5\color{ForestGreen}($\uparrow$12.3\%) \\
+QC 7B w/ G-CLC SFT & 60.0\color{ForestGreen}($\uparrow$8.7\%) &71.4\color{ForestGreen}($\uparrow$15.3\%) \\
\midrule
+QC 14B & 58.1& 65.7 \\
+QC 14B w/ SFT & 58.1\color{gray}($=$0.0\%) & 68.6\color{ForestGreen}($\uparrow$4.4\%) \\
+QC 14B w/ CLC SFT & 59.0\color{ForestGreen}($\uparrow$1.5\%) & 69.5\color{ForestGreen}($\uparrow$5.8\%)\\
+QC 14B w/ G-CLC SFT & 59.0\color{ForestGreen}($\uparrow$1.5\%) & 70.5\color{ForestGreen}($\uparrow$7.3\%) \\
     \bottomrule
    \end{tabular}
    }
    \end{subtable}
    \hfill
    \begin{subtable}[t]{0.48\textwidth}
    \centering
    \caption{Using DS-R1 as edit sketch generation model}
    \resizebox{\linewidth}{!}{
        \begin{tabular}{l|ll}
    \toprule
    Methods & Lazy & Descriptive \\
     \midrule
     +QC 0.5B &21.9 &30.5 \\
     +QC 0.5B w/ SFT &40.0\color{ForestGreen}($\uparrow$82.6\%)&53.3\color{ForestGreen}($\uparrow$74.8\%)\\
     +QC 0.5B w/ CLC SFT &42.9\color{ForestGreen}($\uparrow$95.9\%)&56.2\color{ForestGreen}($\uparrow$84.3\%)\\
      +QC 0.5B w/ G-CLC SFT &47.6\color{ForestGreen}($\uparrow$117.4\%)&60.9\color{ForestGreen}($\uparrow$99.7\%)\\
    \midrule
    +QC 1.5B &34.3 &43.8 \\
    +QC 1.5B w/ SFT &49.5\color{ForestGreen}($\uparrow$44.3\%)&58.1\color{ForestGreen}($\uparrow$32.6\%)\\
 +QC 1.5B w/ CLC SFT &53.3\color{ForestGreen}($\uparrow$55.4\%)&60.0\color{ForestGreen}($\uparrow$37.0\%)\\
 +QC 1.5B w/ G-CLC SFT &56.2\color{ForestGreen}($\uparrow$63.8\%)&62.9\color{ForestGreen}($\uparrow$43.6\%)\\
 \midrule
    +QC 3B &50.5 &60.0 \\
    +QC 3B w/ SFT &57.1\color{ForestGreen}($\uparrow$13.1\%)&66.7\color{ForestGreen}($\uparrow$11.2\%)\\
    +QC 3B w/ CLC SFT &58.1\color{ForestGreen}($\uparrow$15.0\%)&68.6\color{ForestGreen}($\uparrow$14.3\%)\\
    +QC 3B w/ G-CLC SFT&59.0\color{ForestGreen}($\uparrow$16.8\%)&70.5\color{ForestGreen}($\uparrow$17.5\%)\\
    \midrule
    +QC 7B &58.1 &66.7 \\
    +QC 7B w/ SFT&59.0\color{ForestGreen}($\uparrow$1.5\%)&69.5\color{ForestGreen}($\uparrow$4.2\%)\\
    +QC 7B w/ CLC SFT&60.9\color{ForestGreen}($\uparrow$4.8\%)&71.4\color{ForestGreen}($\uparrow$7.0\%)\\
    +QC 7B w/ G-CLC SFT&60.9\color{ForestGreen}($\uparrow$4.8\%)&72.4\color{ForestGreen}($\uparrow$8.5\%)\\
    \midrule
    +QC 14B &61.9&70.5 \\
    +QC 14B w/ SFT &60.9\color{red}($\downarrow$1.6\%)&71.4\color{ForestGreen}($\uparrow$1.3\%)\\
    +QC 14B w/ CLC SFT &61.9\color{gray}($=$0.0\%)&71.4\color{ForestGreen}($\uparrow$1.3\%)\\
    +QC 14B w/ G-CLC SFT &61.9\color{gray}($=$0.0\%)&72.4\color{ForestGreen}($\uparrow$2.7\%)\\
     \bottomrule
    \end{tabular}
    }
    \end{subtable}
    \label{tab:canitedit_train}
    \vspace{-6pt}
\end{table}

In this RQ, we conduct ablation studies to examine how different training strategies for sketch application models affect cascaded code editing performance. We evaluate models trained with different methods in conjunction with DeepSeek-V3 and DeepSeek-R1 for cascaded code editing on the Aider and CanItEdit benchmarks, with results presented in Tables \ref{tab:aider_train} and \ref{tab:canitedit_train}.

Our ablation results demonstrate that all training strategies consistently improve cascaded code editing performance compared to base models. On the Aider benchmark using DeepSeek-V3 as the first-stage edit sketch generation model, we observe that SFT, CLC-SFT, and G-CLC SFT strategies yield average Pass@2 improvements of 29.7\%, 44.9\%, and 60.6\%, respectively, across the five experimented models. This systematic comparison reveals that enhancing sketch application capability alone can yield substantial improvements in overall cascaded code editing performance.

Among the training strategies, our proposed CLC SFT and G-CLC SFT methods demonstrate clear advantages over standard SFT in our ablation study. Specifically, G-CLC SFT shows the most substantial improvements, outperforming standard SFT by remarkable margins across both sketch generation models. When compared to CLC SFT across DeepSeek-V3 and DeepSeek-R1 as edit sketch generation models, G-CLC SFT achieves additional average improvements in Pass@1 and Pass@2 of 6.5\% and 6.6\%, respectively, further validating the effectiveness of our design to improve generalization capabilities.

We observe consistent patterns on the CanItEdit benchmark, where our ablation study reveals that while standard SFT provides baseline improvements of 25.7\%, our proposed CLC SFT and G-CLC SFT methods achieve substantially higher gains of 33.2\% and 39.7\%, respectively, for Pass@1 on Lazy and Descriptive NL requirements. The systematic performance advantages of our proposed training strategies over standard SFT across both benchmarks confirm that specialized training methodologies for sketch application are crucial for optimizing cascaded code editing frameworks.


\subsection{RQ4: Generalization to Other LLMs}

\begin{table}[t]
    \centering
    \caption{Generalization of our framework across various models.}
    \vspace{-6pt}
    \resizebox{\textwidth}{!}{
    \begin{tabular}{l|cc|cc|cc|cc|cc}
    \toprule
    \multirow{2}{*}{Methods} 
    & \multicolumn{2}{c|}{Qwen3 235B} 
    & \multicolumn{2}{c|}{Qwen3 Coder 480B} 
    & \multicolumn{2}{c|}{GPT-OSS-120B} 
    & \multicolumn{2}{c|}{\add{Qwen2.5 Coder 14B}} 
    & \multicolumn{2}{c}{\add{Qwen3 Coder 30B}} \\
    
     & Pass@1 & Pass@2 & Pass@1 & Pass@2 & Pass@1 & Pass@2 & Pass@1 & Pass@2 & Pass@1 & Pass@2 \\
    \midrule
    
    Direct 
    & 24.9 & 55.1 & 29.3 & 55.6 & 15.6 & 53.3 
    & 2.7 & 8.9 & 11.1 & 27.1 \\
    
    + QC 14B 
    & 24.0 & 57.3 & 27.1 & 56.0 & 18.7 & 54.2 
    & 2.7 & 11.1 & 11.6 & 28.0 \\
    
    + QC 14B w/ G-CLC SFT 
    & \textbf{29.3}& \textbf{59.1}& \textbf{32.4} & \textbf{58.7} & \textbf{20.4} & \textbf{55.1}
    & \textbf{4.9} & \textbf{13.3} & \textbf{12.4} & \textbf{30.2} \\
    
    \bottomrule
    \end{tabular}
    }
    \label{tab:general}
    \vspace{-6pt}
\end{table}

We compare the performance of Qwen3-235B, Qwen3-Coder-480B, and GPT-OSS-120B in the cascaded method against direct baselines on the Aider benchmark, as shown in Table \ref{tab:general}. The results indicate that employing Qwen2.5-Coder 14B as the edit sketch application model yields comparable performance across all three models: while Pass@1 achieves comparable performance, Pass@2 increases by 1.5\% on average. Following our data curation and G-CLC SFT enhancements, performance improves further; compared to direct code editing, the average Pass@1 and Pass@2 across the three models rise by 19.7\% and 5.4\%, respectively. This further validates the effectiveness of cascaded code editing and our enhancement methods in generalizing to diverse LLMs.

\add{To assess practical applicability, we conduct additional experiments using Qwen2.5-Coder 14B and Qwen3-Coder 30B \citep{qwen3coder} as edit sketch generation models, paired with fine-tuned smaller models for the application phase. As shown in Table \ref{tab:general}, our cascaded framework consistently improves performance: for Qwen2.5-Coder 14B, Pass@2 increases from 8.9 to 13.3 (+49.4\%), and for Qwen3-Coder 30B, from 27.1 to 30.2 (+11.4\%). However, unlike directly employing frontier large models, the cascaded approach does not yield substantial latency or cost reductions for these deployable-scale models during the application phase, as their base generation overhead is already minimal. These results demonstrate that our framework can also effectively enhance the code editing capabilities of deployable models, making them more practical for real-world applications. }

\section{Discussion}

\subsection{Impact of Direct Code Editing Approaches}

To simulate realistic code editing scenarios, we explore two alternative approaches for direct code editing: (1) search and replace, where the model generates pairs of original and edited code snippets for in-place replacement; and (2) unified diff format, commonly used in Git, where the model generates a patch subsequently merged into the original code. The results on the Aider benchmark are presented in Table \ref{tab:format}.

The results indicate that the whole file and search and replace formats exhibit comparable performance and token usage. For DeepSeek-V3, the search and replace method yields Pass@1 and Pass@2 scores that are lower by 1.3 and 0.4 points, respectively; moreover, for DeepSeek-R1, these metrics are also competitive. Additionally, search and replace consumes slightly fewer tokens than the whole file format, resulting in marginally reduced inference time and cost. In contrast, the unified diff format performs the worst in our experiments, with Pass@1 decreasing by 2.2 and 4.0 points for V3 and R1, respectively, while also incurring the highest token consumption. This suggests that for direct code editing, whole file and search and replace are competitive options. 

\add{\textbf{The ``Context Tax'' in Scoped Editing.} While scoped edits intuitively seem more efficient, our experiments reveal a critical overhead: to ensure unambiguous localization, both unified diff and search and replace must generate \texttt{[Old Code]} + \texttt{[New Code]} pairs. Since \texttt{[Old Code]} serves as a search anchor, models generate extensive surrounding context to disambiguate similar or repeated code patterns. This context requirement—which we term the ``Context Tax''—negates theoretical efficiency gains and explains why unified diff consumes more tokens than whole file generation despite its narrower scope.}

\add{\textbf{How Cascaded Editing Achieves Superior Efficiency and Accuracy.} Our approach bypasses the Context Tax through a fundamentally different division of labor. The Stage 1 Large Model generates a concise edit sketch containing primarily new logic without the burden of extensive matching context (reducing tokens to $\sim$221.5K for DeepSeek-V3). The Stage 2 Small Model then handles context matching and merging efficiently. This decomposition eliminates the heavyweight context generation from the expensive Large Model while offloading precise localization to a specialized, cost-effective component. Additionally, by focusing the Large Model on core logic rather than verbose context, we reduce attention drift in long-context scenarios, leading to higher generation fidelity (92.8\% CodeBLEU) and lower Over-Edit Rate (31.6\% vs. 40.1\%), as shown in Table \ref{tab:local}. Our cascaded framework thus outperforms direct editing across all formats, achieving both superior efficiency and a higher performance ceiling.}

\begin{table}[t]
    \centering
    \caption{Evaluating different edit approaches for direct code editing in the Aider benchmark using DeepSeek-V3 (left) and DeepSeek-R1 (right).}
    \vspace{-6pt}
    \resizebox{\linewidth}{!}{
    \begin{tabular}{c|lllll|lllll}
    \toprule
    Format  & Pass@1 & Pass@2 & \# Tokens & Time & Cost & Pass@1 & Pass@2 & \# Tokens & Time & Cost  \\
    \midrule
     Whole File  & 29.3 & 56.0 & 357.1K & 3.76h & 0.73\$&35.6 & 67.6 & 2.6M & 27.31h & 6.54\$\\
     Search \& Replace & 28.0 & 55.6 & 352.5K & 3.74h& 0.73\$& 34.7 & 65.8 & 2.5M & 26.18h &6.32\$\\
     Unified Diff & 27.1 & 52.4 & 381.2K & 3.93h & 0.75\$&28.4 & 64.0 & 2.8M & 29.74h& 7.01\$\\
     \bottomrule
    \end{tabular}
    }
    \label{tab:format}
    \vspace{-6pt}
\end{table}

\subsection{Effect of Natural Language Sketch Descriptions}

\begin{table}[t]
    \centering
    \caption{The effect of NL sketch description on cascaded code editing using DeepSeek-V3 as edit sketch generation model. \textcolor{ForestGreen}{Green} indicates improvements, \textcolor{red}{red} indicates degradations or increased costs.}
    \vspace{-6pt}
    \resizebox{\textwidth}{!}{%
    \begin{tabular}{l|l|ll|ll|ll|ll}
    \toprule
     \multirow{2}{*}{Model} & \multirow{2}{*}{Methods} & \multicolumn{4}{c|}{w/o description} & \multicolumn{4}{c}{w/ description} \\
     \cline{3-10}
     & & Lazy & Descriptive & Tokens & Time & Lazy & Descriptive & Tokens & Time \\
     \midrule
     \multirow{2}{*}{QC 0.5B} 
     & Original & 20.9 & 28.6 & \multirow{2}{*}{37.5K} & \multirow{2}{*}{0.43h} & 29.5 {\color{ForestGreen}($\uparrow$41.1\%)} & 33.3 {\color{ForestGreen}($\uparrow$16.4\%)} & \multirow{2}{*}{71.4K {\color{red}($\uparrow$90.4\%)}} & \multirow{2}{*}{0.76h {\color{red}($\uparrow$76.7\%)}} \\
     & + G-CLC SFT & 51.4 & 57.1 & & & 53.3 {\color{ForestGreen}($\uparrow$3.7\%)} & 58.1 {\color{ForestGreen}($\uparrow$1.8\%)} & & \\
     \midrule
     
     \multirow{2}{*}{QC 1.5B} 
     & Original & 36.2 & 42.9 & \multirow{2}{*}{37.8K} & \multirow{2}{*}{0.48h} & 43.8 {\color{ForestGreen}($\uparrow$21.0\%)} & 47.6 {\color{ForestGreen}($\uparrow$11.0\%)} & \multirow{2}{*}{72.3K {\color{red}($\uparrow$91.3\%)}} & \multirow{2}{*}{0.82h {\color{red}($\uparrow$70.8\%)}} \\
     & + G-CLC SFT & 53.3 & 63.8 & & & 54.3 {\color{ForestGreen}($\uparrow$1.9\%)} & 62.9 {\color{red}($\downarrow$1.4\%)} & & \\
     \midrule
     
     \multirow{2}{*}{QC 3B} 
     & Original & 48.6 & 53.3 & \multirow{2}{*}{35.8K} & \multirow{2}{*}{0.50h} & 58.1 {\color{ForestGreen}($\uparrow$19.5\%)} & 60.0 {\color{ForestGreen}($\uparrow$12.6\%)} & \multirow{2}{*}{72.6K {\color{red}($\uparrow$102.8\%)}} & \multirow{2}{*}{0.84h {\color{red}($\uparrow$68.0\%)}} \\
     & + G-CLC SFT & 59.0 & 67.6 & & & 59.0 {\color{gray}($=$0.0\%)} & 66.7 {\color{red}($\downarrow$1.3\%)} & & \\
     \midrule

     \multirow{2}{*}{QC 7B} 
     & Original & 55.2 & 61.9 & \multirow{2}{*}{37.9K} & \multirow{2}{*}{0.56h} & 57.1 {\color{ForestGreen}($\uparrow$3.4\%)} & 60.0 {\color{red}($\downarrow$3.1\%)} & \multirow{2}{*}{73.1K {\color{red}($\uparrow$92.9\%)}} & \multirow{2}{*}{0.89h {\color{red}($\uparrow$58.9\%)}} \\
     & + G-CLC SFT & 60.0 & 71.4 & & & 60.0 {\color{gray}($=$0.0\%)} & 71.4 {\color{gray}($=$0.0\%)} & & \\
     \midrule

     \multirow{2}{*}{QC 14B} 
     & Original & 58.1 & 65.7 & \multirow{2}{*}{37.5K} & \multirow{2}{*}{0.62h} & 58.1 {\color{gray}($=$0.0\%)} & 70.5 {\color{ForestGreen}($\uparrow$7.3\%)} & \multirow{2}{*}{73.9K {\color{red}($\uparrow$97.1\%)}} & \multirow{2}{*}{0.94h {\color{red}($\uparrow$51.6\%)}} \\
     & + G-CLC SFT & 59.0 & 70.5 & & & 59.0 {\color{gray}($=$0.0\%)} & 71.4 {\color{ForestGreen}($\uparrow$1.3\%)} & & \\
     \bottomrule
    \end{tabular}
    }
    \label{tab:nl_code}
    \vspace{-6pt}
\end{table}

While our main experiments focus on generating sketches of pure code changes, real-world code modifications sometimes include natural language (NL) annotations that explain the rationale or purpose of the changes. Motivated by this observation, we investigate the impact of incorporating NL elements in generated sketches. Specifically, we extend our edit sketch generation model to produce not only code snippets but also accompanying NL descriptions of the modifications on the CanItEdit benchmark.

Table \ref{tab:nl_code} presents the experimental results. We find that for smaller original instruction models (0.5B–3B parameters), adding NL modification descriptions yields positive effects, with average improvements in Lazy and Descriptive Pass@1 of 27.2\% and 13.3\%, respectively, across these three models. For relatively larger models (7B and 14B), the impact is minimal or even negative, with occasional declines (e.g., a 3.1\% drop in Descriptive Pass@1 for the 7B model). However, incorporating NL descriptions substantially increases the number of tokens generated in the edit sketch generation phase, with an average elevation of 94.9\% in sketch tokens across the five experiments, leading to a corresponding 65.2\% increase in total inference time. This diminishes the efficiency gains of the cascaded approach. In contrast, our trained models exhibit much smaller influences from NL descriptions: for the 0.5B–3B sizes, average impact in Lazy and Descriptive Pass@1 are only 1.9\% and -0.3\%, respectively, while changes for 7B and 14B are also negligible. This indicates that our method substantially reduces dependency on NL descriptions in sketches, maintaining high accuracy while preserving efficiency.

\subsection{\add{Discussion about Localization and Generation}}

\begin{table}[t]
\centering
\caption{Analysis of Localization and Generation Performance.}
\vspace{-8pt}
\label{tab:local}
\resizebox{1\textwidth}{!}{%
\begin{tabular}{llcc}
\toprule
\textbf{Category} & \textbf{Metric} & \textbf{Scoped Direct Editing (S\&R)} & \textbf{Cascaded Code Editing (Ours)} \\
\midrule
\multirow{2}{*}{\textit{Localization}} 
& Over-Edit Rate $\downarrow$ & 40.1\% & \textbf{25.6\%} \\
& Under-Edit Rate $\downarrow$ & 10.5\% & \textbf{8.3\%} \\
\midrule
\textit{Generation} 
& Generation Accuracy $\uparrow$ & 87.5\% & \textbf{92.8\%} \\
\midrule
\multirow{2}{*}{\textit{Stage 2 Fidelity}} 
& Application Over-Edit $\downarrow$ & \multirow{2}{*}{N/A} & 5.2\% \\
& Application Under-Edit $\downarrow$ & & 2.1\% \\
\bottomrule
\end{tabular}%
}
\end{table}

\add{In this section, we evaluate the localization and generation capabilities of Scoped Direct Edit and our method using our human-annotated benchmark. Specifically, Scoped Direct Edit (S\&R) is a straightforward baseline, where the LLM directly outputs both the target location and the corresponding modification in a single pass, with edits applied via pattern-based search-and-replace. However, this paradigm exhibits inherent limitations in long-context scenarios, as the joint optimization of edit localization and content generation within a single inference pass amplifies susceptibility to contextual ambiguity and hallucinated outputs. To comprehensively assess performance, we employ three complementary metrics. For localization accuracy, we measure Over-Edit Rate and Under-Edit Rate: the former quantifies the proportion of modified lines that deviate from the ground truth, capturing unnecessary changes, while the latter measures required modifications that are missing in the final output. For generation quality, we compute the CodeBLEU score between predicted and ground-truth diff blocks. Additionally, for our cascaded framework, we introduce Stage 2 Fidelity metrics that isolate the precision of the edit application step by measuring Application Over-Edit Rate (hallucinated edits not specified in the sketch) and Application Under-Edit Rate (sketch edits omitted during application). The results are presented in Table~\ref{tab:local}.}

\add{From the results, we can observe that cascaded editing significantly reduces localization and generation errors compared to the Scoped Direct Edit baseline, achieving 92.8\% CodeBLEU and a 14.5\% lower over-edit rate. It is worth noting that the relatively high overall over-edit rate does not necessarily signify erroneous modifications, as the model may tend to generate functionally benign edits that align with its stylistic preferences, such as adding or revising comments \citep{wang2025beyond2, cassanocan}. Furthermore, by offloading the implementation phase, the Stage 1 large model avoids context-driven distractions, leading to more focused and accurate edit site identification. Simultaneously, our specialized Stage 2 applicator demonstrates high-fidelity execution with minimal application-level errors (5.2\% over-edit and 2.1\% under-edit). This synergy proves that our cascaded code editing not only prevents the reasoning fatigue of general-purpose models in long-context files but also provides a more reliable mechanism for precise code merging than direct editing.}



\subsection{Threats to Validity}

While our evaluation demonstrates promising results for the cascaded code editing framework, several threats to validity should be acknowledged. First, the experiments are conducted on a limited set of models, such as DeepSeek series and Qwen2.5-Coder variants, and benchmarks, which may not fully represent the diversity of real-world LLMs and code editing scenarios. Thus, further validation across a broader range of models and datasets is necessary to confirm generalizability. Additionally, the performance gains observed in editing the calculated time costs and efficiency improvements in the cascaded approach could be influenced by real-world factors, including varying computational resources, hardware configurations, and network latencies.

\section{Related Work}

\subsection{Code Editing}
Eclipse and Visual Studio integrated rule-based automated code edit tools relying on static analysis for some specific applications such as syntactic correctness. With the development of deep learning, deep neural networks such as RNNs and Transformers have been applied to code editing tasks~\citep{DBLP:conf/kbse/ChakrabortyR21,DBLP:journals/tse/ChakrabortyDAR22,DBLP:journals/tosem/LiLLJHZF23,DBLP:conf/issta/LiuCLHPJYD024}. Recently, LLM significantly advanced this field. InstructCoder~\citep{DBLP:conf/acl/LiHZCXLSH24} fine-tunes Code LLaMA~\citep{codellama} with for general-purpose code editing from nature language instructions. CanItEdit~\citep{cassanocan} is a benchmark to assess LLMs' instructional code editing abilities. Meanwhile, Nam et al.~\citep{DBLP:journals/corr/abs-2504-20196} analyzed IDE telemetry logs from Google, revealing developers' struggles with prompting, and proposed the AutoPrompter tool to help developer edit their code. NextCoder~\citep{aggarwal2025nextcoder} developed a synthetic data pipeline and a robust adaptation SeleKT to train models on code-editing tasks.


\subsection{LLM for Code}
The advent of LLMs has significantly advanced progress in software engineering. Various foundational code LLMs have been developed, such as Code Llama~\citep{roziere2023code}, DeepSeek-Coder~\citep{guo2024deepseek}, and Qwen-Coder~\citep{qwen3coder}. A significant body of research focuses on enhancing these models and applying them to diverse software engineering tasks. To improve their core capabilities, researchers have developed novel training methodologies, such as synthesizing fine-tuning datasets guided by APIs~\citep{li2025api}. Other works improve the code generation process through self-correction; for instance, Self-Debug~\citep{DBLP:conf/iclr/ChenLSZ24} leverages execution feedback for reflection and correction, and ChatUniTest~\citep{DBLP:conf/sigsoft/ChenHZHDY24} uses a generation-validation-repair cycle to fix errors in unit tests. In parallel, LLMs are being applied to a broadening spectrum of tasks. They have been used to decompose complex coding problems into executable plans~\citep{DBLP:journals/pacmse/BairiSKCIPRAS24}, generate proofs for verification~\citep{DBLP:conf/icse/ChakrabortyEBFF25}, create input generators for fuzzing~\citep{zhang2025low}, and even solve software security challenges~\citep{ji2025measuring}. 
\section{Conclusion}

We have presented a cascaded code editing framework that decomposes the task into edit sketch generation and application phases, effectively addressing the effectiveness-efficiency trade-off in LLM-based code editing. To enhance smaller models' sketch application capabilities, we have constructed the first large-scale dataset and have proposed progressive training strategies, including CLC SFT and G-CLC SFT. Our experiments have demonstrated that the cascaded approach achieves superior performance while reducing inference time and cost. This work has established a new paradigm for efficient code editing that leverages the complementary strengths of large and small models, paving the way for more scalable AI-assisted development tools.

\section*{Acknowledgment}
The work described in this paper was supported by  Innovation and Technology Commission of Hong Kong SAR Government (GHP/146/22SZ of the Guangdong-Hong Kong Technology Cooperation Funding Scheme), and Research Grants Council of the Hong Kong Special Administrative Region, China (No. SRFS2425-4S03 of the Senior Research Fellow Scheme). This research is also supported by National Natural Science Foundation of China under project (No. 62472126).

\section*{Data Availability}
The data and source code used in this paper can be accessed in \url{https://github.com/adf1178/cascaced_code_editing}.

\bibliographystyle{ACM-Reference-Format}
\bibliography{sample-base,zjNewFull2509}


\end{document}